\documentclass[a4paper,aps,pre,11pt]{revtex4}

\usepackage{color}
\usepackage{ulem}

\begin{document}

\title{Quasi-Lie Brackets and the Breaking of Time-Translation Symmetry for
Quantum Systems Embedded in Classical Baths}

\author{Alessandro Sergi\footnote{asergi@unime.it}}

\affiliation{
Dipartimento di Scienze Matematiche e Informatiche, Scienze Fisiche e Scienze della Terra,
Università degli Studi di Messina, Contrada Papardo, 98166 Messina, Italy}

\affiliation{Institute of Systems Science,
Durban University of Technology,
P. O. Box 1334, Durban 4000, South Africa}

\affiliation{Istituto Nazionale di Fisica Nucleare, Sez. di Catania,
Catania 95123, Italy}

\author{Gabriel Hanna\footnote{gabriel.hanna@ualberta.ca}}
\affiliation{Department of Chemistry, University of Alberta,
11227 Saskatchewan Driven, Edmonton, {AB}~T6G~2G2, Canada}

\author{Roberto Grimaudo\footnote{roberto.grimaudo01@unipa.it}}

\affiliation{Dipartimento di Fisica e Chimica
dell’Universit\'a di Palermo, 
Via Archirafi 36, I-90123 Palermo, Italy}

\affiliation{Istituto Nazionale di Fisica Nucleare, Sez. di Catania,
Catania 95123, Italy}

\author{Antonino Messina\footnote{antonino.messina@unipa.it}}

\affiliation{Dipartimento di Matematica e Informatica,
Universit\`a degli Studi di Palermo, 
Via Archirafi 34, I-90123~Palermo, Italy}

\affiliation{Istituto Nazionale di Fisica Nucleare, Sez. di Catania,
Catania 95123, Italy}

\begin{abstract}
Many open quantum systems encountered in both natural
and synthetic situations are embedded in
classical-like baths. Often, the bath degrees of freedom may be
represented in terms of canonically conjugate coordinates, but in some cases
they may require a non-canonical or non-Hamiltonian representation.
Herein, we review an approach to the dynamics and statistical mechanics
of quantum subsystems embedded in either non-canonical or
non-Hamiltonian classical-like baths which is based on operator-valued
quasi-probability functions. These
functions typically evolve through the action of quasi-Lie brackets
and their associated Quantum-Classical Liouville Equations.
\end{abstract}

\maketitle


\section{Introduction}

A growing community of physicists is interested in both monitoring 
and controlling the time evolution of 
small numbers of quantum degrees of freedom (DOF) that
are embedded in noisy and uncontrollable environments~\cite{caldeira,weiss,bp}.
A specific case of such a system
is encountered when the environment is classical-like in nature.
This situation is one of fundamental importance because, ultimately,
we and our experimental tools behave classically, at least from a coarse-grained
perspective.
In recent years, we have also witnessed a rising interest in
nano-mechanical, opto-mechanical and other types of hybrid quantum
systems~\cite{nosc,nosc2,nosc3,nosc4,nosc5,nosc6,nosc7,nosc8,nosc9,nosc10,nosc11,nosc12,nosc13,nosc14,nosc15,nosc16,nosc17,nosc18,nosc19,nosc20,nosc21,nosc22,nosc23}.
Such systems often exhibit an interplay between classical and quantum effects, 
allowing them to be modeled by means of hybrid quantum-classical methods.

It has been known for a long time, that the dynamics and statistical mechanics
of a quantum subsystem coupled to classical-like DOF
can be formulated in terms of operator-valued quasi-probability
functions in phase space~\cite{silin,rukhazade-silin,balescu-1968,zhang-balescu,balescu-zhang,osborn}.
For example, the dynamics of nano-mechanical oscillators 
has been previously described by one of the authors in terms of
operator-valued quasi-probability functions~\cite{beck}.
Such functions evolve through
quasi-Lie brackets~\cite{aleksandrov,gerasimenko,boucher-transchen,gerasimenko2,martens-fang,prezhdo-kisil,donoso-martens,donoso-martens2,qc-stat,b3,b4},
which can also be augmented by dissipative terms when the energy
is not conserved~\cite{ray-barbados,langevin-qc}.
When the bath is described by canonically conjugate variables
(and only in this case), a hybrid quantum-classical formalism may be derived.
Starting from a fully quantum representation of the subsystem
and bath DOF, one can perform a partial Wigner transform~\cite{wigner}
(over the bath DOF) and then take its semiclassical limit~\cite{kapra-cicco}.
The resulting equation of motion is commonly known as the
quantum-classical Liouville equation (QCLE)~\cite{donoso-martens3,donoso-martens4,wan-schofield,wan-schofield2,stock,schutte,wan-schofield3,schutte2,schutte3,mqc-rate,schutte4,riga-martens,roman-martens}.
The QCLE has been used to study a wide variety of problems~\cite{thorndyke-micha,mqc-prot,riga-martens2,isotope,h-complex,2spins,spectra,etransf,opto,prot-el,plasmon,prot-el2,pump-probe,vibr,helix}
and a number of in-depth reviews on the basic
formulation of the theory exist~\cite{rev,simu,sergi-theor,rev1,rev2,rev3,rev4,rev5,rev6,rev7,rev8,rev9,rev10,rev11}.
The mathematical structure
underlying the QCLE is dictated by a quasi-Lie bracket~\cite{b3,b4,sergi-spin1,sergi-spin2}.
Quasi Lie brackets are known within the community of
classical molecular dynamics simulators as non-Hamiltonian
brackets~\cite{b1,b2,aspvg}.
Mathematicians have also studied very similar structures known as almost Poisson brackets
or quasi-Lie algebras~\cite{qLie,qLie2,qLie3,qLie4,qLie5}.
It is interesting to note that the quasi-Lie (or non-Hamiltonian) structure 
of the QCLE~\cite{aleksandrov,gerasimenko,
zhang-balescu,balescu-zhang,boucher-transchen,gerasimenko2,martens-fang,
prezhdo-kisil,donoso-martens,donoso-martens2,qc-stat,b3,b4}
has both favorable and unfavorable aspects
associated with it. Because
the antisymmetry of the quasi-Lie bracket ensures energy conservation,
one is able to verify the stability of numerical integration algorithms.
However, because the quasi-Lie algebra is not invariant under time translation,
the initially classical DOF acquire a quantum character
as time flows, implying that one never has a true dynamical
theory of quantum and classical DOF but only an approximated dynamics
of a full quantum system~\cite{caro-salcedo}.
This~is somewhat paradoxical because energy conservation is linked to
time-translation symmetry through the Noether theorem;
nevertheless, quasi-Lie brackets break the time-translation symmetry
of the algebra (which can be seen as a signature of the effect of the
classical bath on the quantum subsystem).

This review deals with situations where the bath DOF are described in terms of
non-canonical coordinates~\cite{non-c,mccauley} or non-Hamiltonian
coordinates~\cite{b1,b2,aspvg}, and situations where
dissipation must be taken explicitly into account~\cite{ray-barbados,langevin-qc}.
In all these cases, we will see that the operator-valued
probability functions will develop new functional dependences
and novel definitions of the quasi-Lie brackets will have to be introduced.
In particular, we will first describe
the case of a classical spin bath~\cite{sergi-spin1,sergi-spin2},
as an example of a bath described by non-canonical
coordinates~\cite{non-c,mccauley}.
It has been shown that for such a bath an off-diagonal~\cite{manini}
open-path~\cite{pati,filipp,englman} geometric phase~\cite{berry,qphases,mead}
enters into the propagation of the quantum-classical dynamics.
We will then describe the case of
a non-Hamiltonian bath, which arises when the bath
coordinates coupled to the quantum subsystem are also coupled
to a large bath (which does not directly interact with the quantum subsystem
and whose detailed dynamics is not of interest).
In such cases, the secondary bath acts
as a thermal reservoir and can be described either by means of stochastic
processes~\cite{gardiner} (e.g., Langevin dynamics~\cite{langevin-qc}),
or by means of non-Hamiltonian fictitious coordinates acting as deterministic
thermostats (e.g., the Nos\'e--Hoover thermostat~\cite{nose,hoover}).
Both Langevin and Nos\'e--Hoover deterministic
time evolutions are examples of non-Hamiltonian dynamics.
However, only Nos\'e--Hoover dynamics is defined solely in terms
of a quasi-Lie bracket~\cite{b3,b4}.
Instead, explicit dissipative dynamics requires that
diffusive terms be added to the bracket.

The quantum-classical equations of motion herein discussed can be
implemented {in silico} using a variety of
simulation algorithms~\cite{sergi-theor,sstp,trotter,asfp,dauasfp,dauasfp2,fmgh,
ddagh,ddagh2,ddagh3,ddagh4,jlgh}.
We will sketch out one such integration algorithm, which unfolds
the quantum-classical dynamics of the operator-valued quasi-probability
function in terms of piecewise-deterministic trajectories evolving
on the adiabatic energy surfaces of the system under study
~\cite{sergi-theor,sstp}.

The structure of this review is as follows. 
In Section~\ref{sec:qcd}, we illustrate the algebraic approach used
to formulate the dynamics of a quantum subsystem embedded in a classical-like
environment with canonically conjugate coordinates.
In Section~\ref{sec:csb},
we show how this formalism can be generalized to the case of a bath described
by non-canonical variables, namely a collection of classical spins. Here,~we~will also show how an off-diagonal open-path geometric phase enters into the
time evolution of the operator-valued quasi-probability function of the system.
In Section~\ref{sec:qcd-sb}, we show how the formalism allows us to also 
treat stochastic classical-like baths undergoing Langevin dynamics. 
Finally,~in~Section~\ref{sec:qcnose},
we shed light on the quasi-Lie algebra established by the quantum-classical 
brackets and show how their antisymmetric structure is exploited to achieve 
thermal control of the bath DOF by means of deterministic thermostats such as 
the Nos\'e--Hoover and Nos\'e--Hoover chain thermostats.
Our conclusions and perspectives are given in Section~\ref{sec:cp}.


\section{Quasi-Lie Brackets and Hybrid Quantum-Classical Systems}
\label{sec:qcd}

Classical and quantum dynamics share the same algebraic
structure~\cite{dirac,balescu}, which is realized
by means of Poisson brackets
in the classical case and commutators in the quantum theory.
Poisson~brackets have a symplectic
structure that is easily represented in matrix 
form~\cite{mccauley,goldstein}.
Both Poisson brackets and commutators define Lie algebras.
In terms of commutators, a Lie algebra possesses the following properties:
\begin{eqnarray}
[\hat\chi_1,\hat\chi_2]&=&
-[\hat\chi_2,\hat\chi_2], \label{antisymmetry} \\
\left[\chi_1\hat\chi_2,\hat\chi_3\right]&=&
\hat\chi_1[\hat\chi_2,\hat\chi_3]+
[\hat\chi_1,\hat\chi_3]\hat\chi_2,
\\
\left[c,\hat\chi_j\right]&=& 0
\label{lie-last},
\end{eqnarray}
where $c$ is a so-called c-number and $\hat \chi_j$, $j=1,2,3$ are
quantum operators.
In order to have a Lie algebra,
together with Equations~(\ref{antisymmetry})--(\ref{lie-last}),
the Jacobi relation must also hold
\begin{equation}
{\cal J}=[\hat\chi_1,[\hat\chi_2,\hat\chi_3]]+
[\hat\chi_3,[\hat\chi_1,\hat\chi_2]]+
[\hat\chi_2,[\hat\chi_3,\hat\chi_1]]
=0 \label{jacobi}.
\end{equation}

The time-translation invariance of the commutator algebra follows
from the Jacobi relation, which therefore states an integrability condition.
If $\hat H$ is not explicitly time-dependent,
the antisymmetry of the commutator~(\ref{antisymmetry}),
arising from the antisymmetry of the symplectic matrix
$\mbox{\boldmath$\cal B$}$, ensures that the energy is a constant of motion:
$d\hat H/dt= i \hat{\cal L}\hat H=0$.
Energy conservation under time-translation
is a fundamental property shared by the algebra 
of Poisson brackets and the algebra of commutators
that is in agreement with Noether theorem.

Now, let us consider a hybrid quantum-classical system,
in which the quantum subsystem,
described by a few canonically conjugate operators
$(\hat q, \hat p)=\hat x$ is embedded in a classical bath
with many DOF, described by many canonically conjugate 
phase space coordinates, $X=(Q, P)$.
We will assume that the Hamiltonian of this hybrid system has the form
\begin{eqnarray}
\hat H_{\rm W}(X)
&=&\frac{P^2}{2M} + \frac{\hat p^2}{2m} + V_{\rm W}(\hat q,Q)
\nonumber\\
&=&\frac{P^2}{2M}+\hat h_{\rm W}(Q) \;,
\label{eq:H_W}
\end{eqnarray}
where $m$ and $M$ are the
masses of the subsystem and bath DOF, respectively, and $V_{\rm W}$ is the potential energy
describing the interactions among the subsystem DOF, among the bath DOF,
and between these two sets of DOF.
The last equality on the right-hand side of Equation~(\ref{eq:H_W}) defines the adiabatic Hamiltonian
$\hat h_{\rm W}(Q)$ of the system. {It has been known for many years that the statistical mechanics of such hybrid quantum-classical systems may be formulated in terms
of an operator-valued quasi-probability function
$\hat W(X,t)$}~\cite{silin,rukhazade-silin,balescu-1968,zhang-balescu,balescu-zhang,osborn}.
Specifically, the statistical average of hybrid quantum-classical operators,
representing a dynamical property of the system, may be
calculated according to
\begin{equation}
\langle\hat\chi\rangle(t)={\rm Tr'}\int dX
\hat W(X,t)\hat\chi_{\rm W}(X) \;,
\label{eq:TrW}
\end{equation}
where ${\rm Tr}'$ denotes the partial trace involving a complete set of
states of the quantum subsystem.

The operator-valued quasi-probability function in phase space
evolves according to 
\begin{equation}
\frac{\partial}{\partial t}\hat W(X,t)
=-\frac{i}{\hbar}
\left[\begin{array}{cc} \hat H_{\rm W} & \hat W(X,t) \end{array}\right]
\mbox{\boldmath$\cal D$}
\left[\begin{array}{c} \hat H_{\rm W} \\ \hat W(X,t) \end{array}\right]
=-\frac{i}{\hbar}\left[\hat H_{\rm W}
,\hat W(X,T)\right]_{\mbox{\boldmath$\cal D$}}
=-i\hat{\cal L}^{\mbox{\boldmath$\cal D$}}\hat\chi\;,
\label{eq:qclm}
\end{equation}
where $\mbox{\boldmath$\cal D$}$ is an antisymmetric matrix super-operator
defined by
\begin{equation}
\mbox{\boldmath$\cal D$}=\left[\begin{array}{cc} 0& 1-
\frac{\overleftarrow{\mbox{\boldmath$\nabla$}}
\mbox{\boldmath${\cal B}$}
\overrightarrow{\mbox{\boldmath$\nabla$}}}
{2i\hbar^{-1}}\\
-\left(1-
\frac{\overleftarrow{\mbox{\boldmath$\nabla$}}
\mbox{\boldmath${\cal B}$}
\overrightarrow{\mbox{\boldmath$\nabla$}}}
{2i\hbar^{-1}}\right) & 0\end{array}\right]\;,
\label{eq:D_matrix}
\end{equation}
with
$\mbox{\boldmath$\nabla$}=(\partial/\partial Q,\partial/\partial P)
=\partial/\partial X$,
and
\begin{equation}
\overleftarrow{\mbox{\boldmath$\nabla$}}\mbox{\boldmath${\cal B}$}\overrightarrow{\mbox{\boldmath$\nabla$}}
=\sum_{I,J=1}^{2N}
\overleftarrow{\nabla}_I
{\cal B}_{IJ}\overrightarrow{\nabla}_J
\label{eq:symplectic-bracket}
\end{equation}
denotes the Poisson bracket operator.
The last equality on the right-hand side of Equation~(\ref{eq:qclm}) defines
the quantum-classical Liouville operator $i\hat{\cal L}^{\mbox{\boldmath$\cal D$}}$.
Equation~(\ref{eq:qclm}) is the QCLE~\cite{donoso-martens3,donoso-martens4,wan-schofield,wan-schofield2,stock,schutte,wan-schofield3,schutte2,schutte3,mqc-rate,schutte4,riga-martens,roman-martens}
of the system.

The QCLE in Equation~(\ref{eq:qclm})
is founded upon a quasi-Lie bracket, which we may write explicitly as
\begin{equation}
[\hat\chi_1(X),\hat\chi_2(X)]_{\mbox{\boldmath$\cal D$}}
=
\left[\begin{array}{cc}\hat\chi_1(X) & \hat\chi_2(X)\end{array}\right]
\mbox{\boldmath$\cal D$}
\left[\begin{array}{c}\hat\chi_1(X) \\ \hat\chi_2(X)\end{array}\right]\;,
\label{eq:gen-quantum-algebra}
\end{equation}
where $\mbox{\boldmath$\cal D$}$ is the
antisymmetric matrix operator defined in Equation~(\ref{eq:D_matrix}).
However, in contrast to the Lie brackets of quantum and
classical mechanics, the quasi-Lie bracket defined in 
Equation~(\ref{eq:gen-quantum-algebra})
violates the Jacobi relation~(\ref{jacobi}):
\begin{equation} \small
{\cal J}_{\mbox{\boldmath$\cal D$}}
=\left[\hat\chi_1(X),\left[\hat\chi_2(X),\hat\chi_3(X)
\right]_{\mbox{\boldmath$\cal D$}}
\right]_{\mbox{\boldmath$\cal D$}}
+\left[\hat\chi_3(X),\left[\hat\chi_1(X),\hat\chi_2(X)
\right]_{\mbox{\boldmath$\cal D$}}
\right]_{\mbox{\boldmath$\cal D$}}
+\left[\hat\chi_2(X),\left[\hat\chi_3(X),\hat\chi_1(X)
\right]_{\mbox{\boldmath$\cal D$}}
\right]_{\mbox{\boldmath$\cal D$}} \neq 0
\;.\label{qc-jacobi}
\end{equation}

The failure of the Jacobi implies that the algebra of quasi-Lie brackets
is not invariant under time-translation.
For example, it can be generally proven that
\begin{eqnarray} 
e^{it{\cal L}^{\mbox{\boldmath$\cal D$}}}\left[\hat\chi_1(X,0),
\hat\chi_2(X)\right]
\neq
\left[e^{it{\cal L}^{\mbox{\boldmath$\cal D$}}}\hat\chi_1(X),
e^{it{\cal L}^{\mbox{\boldmath$\cal D$}}}\hat\chi_2(X)\right] \;.
\end{eqnarray}
On the other hand, the quasi-Lie bracket conserves the energy
$e^{it{\cal L}^{\mbox{\boldmath$\cal D$}}}\hat H {_{\rm W}}(X)
= \hat H{_{\rm W}}(X)$. {Hence, the~dynamics generated by the QCLE displays energy conservation
and lack of time-translation invariance of the bracket algebra. 
The situation is surprising because
one does not expect a broken time-translation invariance symmetry
in an isolated system.
However, while a total hybrid quantum-classical system is closed
from the point of view of energy conservation,
the quasi-Lie bracket describes the irreversible transfer of quantum
information from the subsystem to the classical DOF, which acquire
a quantum character as the time flows.
In this sense,} one can heuristically argue that
the lack of time-translation invariance or the algebra is a mere
consequence of the open dynamics of the quantum subsystem.

\subsection{Derivation of the QCLE through a Partial Wigner Transform}

When the bath DOF are described by canonically conjugate
variables (and only in this case), the~hybrid quantum
can be derived by performing a partial Wigner transform
of the quantum Liouville equation (QLE)
over the bath DOF and taking a semiclassical limit of the resulting equations.
To this end, let us consider the fully quantum counterpart to the Hamiltonian in
Equation~(\ref{eq:H_W}):
\begin{equation}
\hat H = \frac{\hat P^2}{2M} + \frac{\hat p^2}{2m} + V(\hat q,\hat Q)\;.
\label{eq:H}
\end{equation}

The quantum statistical state of the system is described by 
the density matrix (or statistical operator)
$\hat \rho(t)$.
The time dependence of the density matrix is dictated by the QLE:
\begin{eqnarray}
\frac{d}{dt}\hat\rho(t) &=& -\frac{i}{\hbar}\left[\hat H,\hat\rho(t)\right]
=-\frac{i}{\hbar}\left[\begin{array}{cc}\hat\rho & \hat H\end{array}\right]
\mbox{\boldmath$\mathcal B$}
\left[\begin{array}{c}\hat\rho \\ \hat H\end{array}\right]
\;,
\label{eq:qle}
\end{eqnarray}
where $[..., ...]$ denotes the commutator, and
$\mbox{\boldmath${\cal B}$}$ is the symplectic matrix~\cite{mccauley,goldstein}:
\begin{equation}
\mbox{\boldmath$\cal B$}
= \left[ \begin{array}{cc} 0 & 1 \\ -1 & 0 \end{array} \right]
\;.
\label{eq:B_matrix}
\end{equation}

The average of an operator $\hat\chi$ defined on the same Hilbert space
of the system is calculated by
\begin{equation}
\langle\hat\chi\rangle(t) = {\rm Tr}\left(\hat\rho(t)\hat\chi\right) \;,
\label{eq:Trq}
\end{equation}
where ${\rm Tr}$ denotes the trace operation.
Now, in order to derive a classical-like description of the bath,
one introduces the partial Wigner transform of the density matrix
$\hat\rho$ over the $\hat X$'s:
\begin{eqnarray}
\hat W(X,t) &=& \frac{1}{2\pi\hbar} \int dZ e^{iP\cdot Z/\hbar}
\langle Q-\frac{Z}{2}\arrowvert\hat\rho(t)
\arrowvert Q+\frac{Z}{2}\rangle\;.
\label{eq:partWtrho}
\end{eqnarray}

The symbol 
$\hat W$ denotes an operator-valued Wigner function (also known as the 
partially-Wigner transformed density matrix),
which is both an operator in the Hilbert space of the $\hat q$'s
and a function of the bath coordinates $X$.
The partial Wigner transform  of an arbitrary operator
$\hat\chi$ is analogously given~by
\begin{eqnarray}
\hat\chi_{\rm W}(X) &=& \int dZe^{iP\cdot Z/\hbar}
\langle Q-\frac{Z}{2}\arrowvert
\hat\chi\arrowvert Q+\frac{Z}{2}\rangle \;.
\label{eq:partWtchi}
\end{eqnarray}

Taking the partial Wigner transform of Equation~(\ref{eq:Trq})
leads to the expression for the average of $\hat\chi$
given in Equation~(\ref{eq:TrW}).
The partial Wigner transform of the Hamiltonian in
Equation~(\ref{eq:H}) is given in Equation~(\ref{eq:H_W}).

Upon taking the partial Wigner transform of the QLE, Equation~(\ref{eq:qle}),
and truncating the resulting equation after first order in $\hbar$,
one arrives at the QCLE
\begin{eqnarray}
\frac{\partial}{\partial t}\hat W(X,t)
&=&-\frac{i}{\hbar}
\left[\hat H_{\rm W},\hat W(X,t)\right]
+ 
\frac{1}{2} \hat H_{\rm W} \overleftarrow{\mbox{\boldmath$\nabla$}}
\mbox{\boldmath${\cal B}$}
\overrightarrow{\mbox{\boldmath$\nabla$}}\hat W(X,t)
-
\frac{1}{2} \hat W(X,t)
\overleftarrow{\mbox{\boldmath$\nabla$}}
\mbox{\boldmath${\cal B}$}
\overrightarrow{\mbox{\boldmath$\nabla$}}\hat H_{\rm W}
\nonumber\\
&=&
-i{\cal L}\hat W(X,t) \;,
\label{eq:qcle}
\end{eqnarray}
where the last equality defines the quantum Liouville
operator $i{\cal L}=(i/\hbar)[\hat H_{\rm W},\cdot] - (1/2)
(\hat H_{\rm W} \overleftarrow{\mbox{\boldmath$\nabla$}}
\mbox{\boldmath${\cal B}$}
\overrightarrow{\mbox{\boldmath$\nabla$}} \cdot)
+ (1/2)
( \cdot \overleftarrow{\mbox{\boldmath$\nabla$}}
\mbox{\boldmath${\cal B}$}
\overrightarrow{\mbox{\boldmath$\nabla$}} \hat H_{\rm W})$.
To arrive at Equation~(\ref{eq:qcle}),
we have used the partial Wigner transform of a product 
of operators,
\begin{eqnarray}
\left(\hat\chi_1\hat\chi_2\right)_{\rm W}(X)
&=&\hat\chi_{1,\rm W}(X)
e^{\frac{i\hbar}{2} \overleftarrow{\mbox{\boldmath$\nabla$}}
\mbox{\boldmath${\cal B}$}
\overrightarrow{\mbox{\boldmath$\nabla$}}}\hat\chi_{2,\rm W}(X) \;,
\end{eqnarray}
and truncated the exponential after first order in $\hbar$,
{i.e.},
\begin{equation}
e^{\frac{i\hbar}{2} \overleftarrow{\mbox{\boldmath$\nabla$}}
\mbox{\boldmath${\cal B}$}
\overrightarrow{\mbox{\boldmath$\nabla$}}}
\approx
1+ \frac{i\hbar}{2} \overleftarrow{\mbox{\boldmath$\nabla$}}
\mbox{\boldmath${\cal B}$}
\overrightarrow{\mbox{\boldmath$\nabla$}} \;.
\label{eq:linexp}
\end{equation}

It should be noted that Equation~(\ref{eq:linexp}) is exact for
Hamiltonians with quadratic bath terms and bilinear coupling between
the $\hat x$ and $X$ DOF.
In Ref.~\cite{kapra-cicco}, it is shown how the linear expansion
can be performed in terms of the parameter $\mu=\sqrt{m/M}$,
which is small in cases where the bath DOF
are much more massive than those of the subsystem.
Equation~(\ref{eq:qcle}) is exactly equivalent to Equation~(\ref{eq:qclm}).

\subsection{Integration Algorithm}

A number of algorithms, which depend on the basis representation,
exist for approximately solving the QCLE~\cite{wan-schofield,wan-schofield2,wan-schofield3,
schutte2,schutte3,roman-martens,sergi-theor,sstp,trotter,asfp,dauasfp,dauasfp2,fmgh,
ddagh,ddagh2,ddagh3,ddagh4,jlgh}.
Herein, we illustrate the so-called Sequential Short-Time Propagation (SSTP)
algorithm~\cite{sergi-theor,sstp},
which offers a good compromise between accuracy
and simplicity of implementation.
The SSTP algorithm is based on the representation of the QCLE
in the adiabatic basis, which is defined by the eigenvalue equation
\begin{equation}
\hat h_{\rm W}|\alpha;Q\rangle = E_\alpha(Q)|\alpha;Q\rangle \;.
\label{eq:adia_states}
\end{equation}
The representation of the QCLE in the adiabatic basis is sketched in
Appendix~\ref{sec:adbas_rep}.
In the adiabatic basis, the QCLE is given by Equation~(\ref{eq:qcle_adbas})
and the quantum-classical Liouville super-operator matrix
elements are given in Equation~(\ref{eq:liouville}).

To derive the SSTP algorithm, we divide the time interval $t$ into $n$ equal
small steps $\tau=t/n$. If~one is able to calculate the propagation
over a single $\tau$, the dynamics over the whole interval can be reconstructed
by sequential iteration of the procedure.
Let us then consider the quantum-classical propagator
over a small step $\tau$
for the matrix elements of the operator-valued quasi-probability function
$\hat W(X)$ in the adiabatic basis.
Such a propagator is written as
\begin{equation}
\left(e^{-i\tau{\cal L}}\right)_{\alpha\alpha,\beta\beta'}
\approx
\delta_{\alpha\beta} \delta_{\alpha'\beta'}
e^{-i\int_0^\tau ds \omega_{\alpha\alpha'}(s)}
e^{-i\tau L_{\alpha\alpha'}}
\left(1+\tau {\cal T}_{\alpha\alpha',\beta\beta'}\right)
\;.
\label{eq:prop_sstp}
\end{equation}
On the right-hand side of Equation~(\ref{eq:prop_sstp}),
we have introduced $\omega_{\alpha\alpha'}$, the Bohr frequency
defined in Equation~(\ref{eq:Bohr-omega}), $iL_{\alpha\alpha'}$ is
a classical-like Liouville operator,
defined in Equation~(\ref{eq:classical-like_L}),
and ${\cal T}_{\alpha\alpha',\beta\beta'}$ is the transition operator
defined in Equation~(\ref{eq:T_aa'bb'}).
The SSTP dynamics of the matrix elements of $\hat W(X,t)$ is given by
\begin{equation}
W_{\alpha\alpha}(X,\tau)
=
\sum_{\beta\beta'}
\delta_{\alpha\beta} \delta_{\alpha'\beta'}
e^{-i\int_0^\tau ds \omega_{\alpha\alpha'}(s)}
e^{-i\tau L_{\alpha\alpha'}}
\left(1+\tau {\cal T}_{\alpha\alpha',\beta\beta'}\right)
W_{\beta\beta}(X) \;.
\label{eq:prop_sstp_W}
\end{equation}
When $\tau$ is infinitesimal, the right-hand side
of Equations~(\ref{eq:prop_sstp}) and ~(\ref{eq:prop_sstp_W}), 
become essentially equal to the left-hand side, as can be seen from the
Dyson identity~\cite{sstp}.

The transition operator is purely off-diagonal. Its action generates
quantum transitions in the subsystems and changes
the bath momenta accordingly. Upon setting the transition operator
to zero, we obtain an adiabatic expression for the propagator.
If the non-adiabatic effects are not too strong, they may be
treated in a perturbative fashion by sampling
the action of the transition operator in a stochastic fashion.
Typically, researchers have used~ {\cite{b4,mqc-prot,isotope,h-complex,
2spins,spectra,etransf,opto,prot-el,plasmon,prot-el2,pump-probe,vibr,helix,
sergi-theor,rev4,rev7,rev9,rev10,rev11,sergi-spin1,sergi-spin2,sstp,fmgh,
ddagh,ddagh2,ddagh3,ddagh4}}
the following expressions for the probabilities of
making a transition (jump) and not-making a transition, respectively:
\begin{eqnarray}
{\cal P}_{\rm J} &=& \frac{\vert \tau \frac{P}{M}\cdot d_{\alpha\beta}\vert }
{1 + \vert \tau \frac{P}{M}\cdot d_{\alpha\beta}\vert }\;,
\\
Q_{\rm NO-J} &=& \frac{ 1}
{1 + \vert \tau \frac{P}{M}\cdot d_{\alpha\beta}\vert }\;.
\end{eqnarray}
Another important technical ingredient of the algorithm
is the approximation of the transition operator in Equation~(\ref{eq:T_aa'bb'})
with its momentum-jump form:
\begin{eqnarray}
{\cal T}_{\alpha\alpha',\beta\beta'}^{\rm MJ}
&=&
\delta_{\alpha'\beta'}
\frac{P}{M}\cdot d_{\alpha\beta}
e^{(E_\alpha-E_\beta)M\partial/\partial(P\cdot\hat d_{\alpha\beta})^2}
+
\delta_{\alpha\beta}
\frac{P}{M}\cdot d_{\alpha'\beta'}^*
e^{(E_\alpha'-E_\beta')M\partial/\partial(P\cdot\hat d_{\alpha'\beta'}^*)^2}
\label{eq:T_aa'bb'_MJ}
\;,
\end{eqnarray}
where $\hat d_{\alpha\beta}$ is the normalized coupling vector.
Within the momentum-jump approximation~\cite{simu,sergi-theor},
the action of the transition operator on the bath momenta can be easily
obtained in closed form:
\begin{eqnarray}
e^{(E_\alpha-E_\beta)M\partial/\partial(P\cdot\hat d_{\alpha\beta})^2} P
&=&
P - P\left(P\cdot \hat d_{\alpha\beta}\right)
+\hat d_{\alpha\beta}
\sqrt{\left(P\cdot\hat d_{\alpha\beta}\right)^2 
+ M\left(E_\alpha-E_\beta\right)}
\;.
\end{eqnarray}

Considering Equations~(\ref{eq:TrW}) and~(\ref{eq:prop_sstp_W}),
together with its SSTP implementation just described,
one~can see that the solution of the QCLE can be obtained
from an ensemble of classical-like trajectories,
where~each trajectory
(whose initial conditions arise from a Monte Carlo sampling~\cite{frenkel} of the $X$'s),
involves deterministic evolution segments on a given
adiabatic energy surfaces interspersed with stochastic quantum transitions,
caused by the momentum-jump operator in Equation~(\ref{eq:T_aa'bb'_MJ}).

The SSTP algorithm~\cite{sergi-theor,sstp} maps the calculation of
averages through the QCLE~(\ref{eq:qcle}) onto a stochastic process.
It is a hybrid Molecular Dynamics/Monte Carlo procedure suffering from two main problems.
The~first is given by the momentum-jump approximation, which is not valid
in general. One can avoid this approximation by devising different
integration schemes, but~usually at the expense of other approximations~\cite{jlgh}.
The~second problem is not just associated with the SSTP algorithm,
but it is common to all Monte Carlo approaches to the calculation
of quantum averages: the~{\it infamous} sign-problem.
The sign-problem is one of the major unsolved problems in the physics of
quantum systems. Within the SSTP algorithm, it~manifests itself
both through the oscillating phase factors associated with the propagation
on mean-energy surfaces and through the accumulation of fluctuating weights
associated with the Monte Carlo sampling of the quantum transitions.
In~practice, upon~analyzing the results obtained by means of this
algorithm~\mbox{\cite{b4,mqc-prot,isotope,h-complex,2spins,spectra,etransf,opto,prot-el,plasmon,prot-el2,pump-probe,vibr,helix,sergi-theor,rev4,rev7,rev9,rev10,rev11,sergi-spin1,sergi-spin2,sstp,asfp,dauasfp,dauasfp2,fmgh,ddagh,ddagh2,ddagh3,ddagh4}},
we can conclude that the more quantum is the character of the bath the greater is
the error in the calculation of the~averages.

The mapping of the calculation of averages via the SSTP algorithm
onto a stochastic process is reminiscent of the approach to open quantum system dynamics
provided by the Stochastic Liouville Equation (SLE)~\cite{sle1,sle2,sle3,sle4}.
However, in contrast to the SLE, the QCLE is a deterministic equation
that explicitly takes into account all the DOF
of the system without approximating the memory of the total hybrid
quantum-classical system. The stochastic process only enters through
the specific hybrid Molecular Dynamics/Monte Carlo implementation 
provided by the SSTP algorithm. Indeed, a recently proposed scheme of integration~\cite{jlgh}
does not involve any stochastic process~whatsoever.


\section{Classical Spin Baths}
\label{sec:csb}

 {
Contrary to what some books in quantum mechanics state
(in the authors's knowledge, an~exception is Schulman's book~\cite{schulman}),
the concept of spin can be defined in an entirely classical
way~\mbox{\cite{schulman,cartan,hladik,carmeli,barut}}.
In practice, spinors provide a more fundamental representation
of the rotation group than that given by tensors~\cite{schulman,cartan,hladik,carmeli,barut}.
Hence, one can think of a collection, {e.g.}, a bath, of DOF comprising
classical spinors (or, for brevity, spins): a classical spin-bath.
An example of a classical spin baths is given by
the Classical Heisenberg Model~\cite{huang}, whose Hamiltonian is
\begin{equation}
H_{CHS}=\sum_{a=x,y,z}\sum_{I,J}^N S_a^I {\cal C}_{IJ}^a S_a^J \;,
\end{equation}
where ${\bf S}^I$ are $N$ classical vectors obeying the constraint 
\begin{equation}
\left(S_x^I\right)^2 + \left(S_y^I\right)^2 + \left(S_z^I\right)^2 =1 \;,
\end{equation}
for $I=1,...,N$, and the ${\cal C}_{IJ}^a$ are coupling  constants.
}
 {However, since the generalization to baths with many
spins is straightforward, in the following, we will illustrate
the theory using a bath comprising a single classical spin}.
Consider a classical spin vector $\bf S$, with components $S_a$, 
$a=x,y,z$, and~Hamiltonian $H^{\rm S}({\bf S})$.
Let us define the spin gradient as
$\nabla^{\bf S}=\partial/\partial{\bf S}$,
which in terms of the spin components is written as
$\nabla_a^S=\partial/\partial S_a$, with $a=x,y,z$.
The equations of motion of the spin are then written as
\begin{equation}
\dot {\bf S}= {\cal B}^{\bf S} \nabla^{\bf S} H_{\rm S}\;,
\label{eq:eqofm}
\end{equation}
where
\begin{equation}
\mbox{\boldmath $\cal B$}^{\bf S}
=\left[\begin{array}{ccc} 0   &  S_z & -S_y \\ 
                         -S_z &  0   &  S_x \\
                          S_y & -S_x &  0 \end{array}\right] 
\;.
\label{eq:bmat}
\end{equation}

One can also adopt the compact form
${\cal B}_{ab}^{\bf S}=\sum_{c=x,y,z}\epsilon_{abc}S_c$ and $a,b=x,y,z$
of the antisymmetric matrix $\mbox{\boldmath $\cal B$}^{\bf S}$,
where $\epsilon_{abc}$ is the Levi--Civita pseudo-tensor.
The Casimir $C_2={\bf S}\cdot {\bf S}$
is preserved by the equations of motion~(\ref{eq:eqofm}),
independently of the form of the spin Hamiltonian $H^{\rm S}({\bf S})$.
In addition, the~dynamics has a zero phase space compressibility
$\kappa^{\rm S}=\nabla{\bf S}\cdot \dot {\bf S} =0$.
The classical phase space flow of the spin
is defined through the non-canonical bracket
\begin{equation}
\sum_{a,b} A({\bf S})\overleftarrow{\nabla}^{\bf S}_a{\cal B}_{ab}^{\rm S}
\overrightarrow{\nabla}^{\bf S}_b B({\bf S})
=
A({\bf S})\overleftarrow{\nabla}^{\bf S} \mbox{\boldmath$\cal B$}^{\bf S}
\overrightarrow{\nabla}^{\bf S}B({\bf S})\;,
\label{eq:ncan-brack}
\end{equation}
where $A=A({\bf S})$ and $B=B({\bf S})$ are arbitrary functions
of the spin DOF.

Consider now the hybrid quantum-classical Hamiltonian
of a quantum subsystem coupled to the classical spin
\begin{eqnarray}
\hat{\cal H}({\bf S})&=&\hat{H}(\{\hat{\chi}\})
+V_{\rm C}(\{\hat{\chi}\},{\bf S}) +H^{\bf S}({\bf S})
\nonumber\\
&=&
\hat h_{\bf S}({\bf S})+H^{\bf S}({\bf S})\;,
\label{eq:tot-hamS}
\end{eqnarray}
describing a quantum subsystem in terms of the Hamiltonian
$\hat H(\{\hat{\chi}\})$, depending on the operators $\{\hat\chi\}$,
$V(\{\hat\chi\},{\bf S})$ is the subsystem-spin interaction potential, 
and the second line of the equation defines the adiabatic Hamiltonian
$\hat h_{\bf S}$.
The quantum-classical dynamics of the operator-valued quasi-probability 
function (defined in the spinor space of the total system),
$\hat W^{\bf S}({\bf S},t)$, is dictated
by the spin-bath QCLE~\cite{sergi-spin1,sergi-spin2}
\begin{eqnarray}
\frac{\partial}{\partial t}\hat W^{\bf S}({\bf S},t)
&=&-\frac{i}{\hbar}\left[\begin{array}{cc} \hat{\cal H}({\bf S}) & 
\hat W^{\bf S}({\bf S},t)\end{array}\right]
\mbox{\boldmath$\cal D$}^{\rm S}
\left[\begin{array}{c}\hat{\cal H}({\bf S})\\ 
\hat W^{\bf S}({\bf S},t)\end{array}\right]
\;\nonumber\\
&=&-\frac{i}{\hbar}\left[\hat{\cal H}({\bf S}),\hat W^{\bf S}({\bf S},t)
\right]_{\mbox{\tiny \boldmath$\cal D$}^{\rm S}},
\label{eq:pWS}
\end{eqnarray}
where
\begin{eqnarray}
\mbox{\boldmath$\cal D$}^{\rm S}
&=&
\left[\begin{array}{cc} 0 & 1+\frac{i\hbar}{2}
\overleftarrow{\nabla}\mbox{\boldmath$\cal B$}^{\rm S}
\overrightarrow{\nabla}
\\
-1-\frac{i\hbar}{2}
\overleftarrow{\nabla}\mbox{\boldmath$\cal B$}^{\rm S}
\overrightarrow{\nabla}
& 0\end{array}\right]\;.
\end{eqnarray}

We next set out to represent Equation~(\ref{eq:pWS})
in the adiabatic basis $|\alpha;{\bf S}\rangle$
defined by the eigenvalue~equation
\begin{equation}
\hat h_{\bf S}({\bf S})|\alpha;{\bf S}\rangle =E_{\alpha}({\bf S})|\alpha;{\bf S}\rangle\;.
\label{eq:adbasS}
\end{equation}
It should be noted that, in contrast to the case of canonically conjugate 
phase space coordinates which depends only on the positions $Q$ and not on the
conjugate momenta $P$, this adiabatic basis depends on all the non-canonical
spin coordinates $\bf S$.
In this basis, Equation~(\ref{eq:pWS}) becomes
\begin{eqnarray}
\frac{\partial}{\partial t}\hat W^{\bf S}_{\alpha\alpha'}
&=&
-i\omega_{\alpha\alpha'} W^{\bf S}_{\alpha\alpha'}
- H^{\rm S}\overleftarrow{\nabla}^{\bf S}
\mbox{\boldmath$\cal B$}^{\bf S}
\langle\alpha|\overrightarrow{\nabla}^{\bf S}\hat W^{\bf S}|\alpha'\rangle
\nonumber\\
&+&\frac{1}{2}
\langle\alpha|\hat h_{\bf S}\overleftarrow{\nabla}^{\bf S}
\mbox{\boldmath$\cal B$}^{\bf S}
\overrightarrow{\nabla}^{\bf S}\hat W^{\bf S}|\alpha'\rangle
-\frac{1}{2}
\langle\alpha|\hat W^{\bf S}\overleftarrow{\nabla}^{\rm S}
\mbox{\boldmath$\cal B$}^{\rm S}\overrightarrow{\nabla}^{\bf S}
\hat h_{\bf S}|\alpha'\rangle
\;,
\end{eqnarray}
where $\omega_{\alpha\alpha'}=E_\alpha({\bf S}) - E_{\alpha'}({\bf S})/\hbar$
is the Bohr frequency.
Defining the spin coupling vector
\begin{eqnarray}
d^{\bf S}_{\alpha\alpha'}
&=&
\langle\alpha;{\bf S}|\overrightarrow{\nabla}^{\bf S}|\alpha';{\bf S}\rangle,
\;
\end{eqnarray}
one finds the two identities
\begin{eqnarray}
\langle\alpha;{\bf S}|\left(\overrightarrow{\nabla}^{\bf S}
\hat W^{\bf S}({\bf S})\right)|\alpha';{\bf S}\rangle
&=&
\overrightarrow{\nabla}^{\bf S}W^{\bf S}_{\alpha\alpha'}({\bf S})
+\sum_\beta d^{\bf S}_{\alpha\beta}W^{\bf S}_{\beta\alpha'}({\bf S})
-\sum_{\beta'}W^{\bf S}_{\alpha\beta'}({\bf S}) d^{\bf S}_{\beta'\alpha'}
\label{eq:drho}
\\
\langle\alpha;{\bf S}|\left(\overrightarrow{\nabla}^{\bf S}
\hat{h}_{\bf S}({\bf S})\right) |\alpha';{\bf S}\rangle
&=&
\overrightarrow{\nabla}^{\bf S} h_{\bf S}^{\alpha\alpha'}
-\Delta E_{\alpha\alpha'} d^{\bf S}_{\alpha\alpha'}
\label{eq:dh}
\end{eqnarray}
where $\Delta E_{\alpha\alpha'}=E_{\alpha}-E_{\alpha'}$.
Using Equations~(\ref{eq:drho}) and~(\ref{eq:dh}),
the spin-bath QCLE may be rewritten as
\begin{eqnarray}
\frac{\partial}{\partial t}W^{\rm S}_{\alpha\alpha'}({\bf S},t)
&=&-\sum_{\beta\beta'}\left(i\omega_{\alpha\alpha'}\delta_{\alpha\beta}\delta_{\alpha\alpha'}
+iL_{\alpha\alpha'}\delta_{\alpha\beta}\delta_{\alpha\alpha'}
+{\cal T}^{\rm S}_{\alpha\alpha',\beta\beta'}
+{\cal S}_{\alpha\alpha',\beta\beta'}\right)
W^{\rm S}_{\beta\beta'}({\rm S},t)\;,
\label{eq:qcsd-dyna}
\end{eqnarray}
where we have defined the classical-like spin-Liouville operator
\begin{eqnarray}
iL_{\alpha\alpha'}&=&
H_{\rm S}
\overleftarrow{\nabla}^{\rm S}
\mbox{\boldmath$\cal B$}^{\rm S}
\overrightarrow{\nabla}^{\rm S}
+
\frac{1}{2} E_{\alpha'} \overleftarrow{\nabla}^{\rm S}
\mbox{\boldmath$\cal B$}^{\rm S} \overrightarrow{\nabla}^{\rm S}
+
\frac{1}{2} E_{\alpha} \overleftarrow{\nabla}^{\rm S}
\mbox{\boldmath$\cal B$}^{\rm S} \overrightarrow{\nabla}^{\rm S}
\nonumber\\
&=&
\left( \mbox{\boldmath$\cal B$}^{\rm S} \overrightarrow{\nabla}^{\rm S}
H_{\alpha\alpha'}^{\rm S} \right) \cdot
\overrightarrow{\nabla}^{\rm S}
\;,
\label{eq:Lspin}
\end{eqnarray}
with the average adiabatic Hamiltonian
\begin{equation}
H_{\alpha\alpha'}^{\rm S}
=
H_{\rm S}+\frac{1}{2}\left(E_{\alpha}+E_{\alpha'}\right)
\;.
\end{equation}

The transition operator for the spin bath is given by
\begin{eqnarray}\begin{array}{lll}
{\cal T}^{\bf S}_{\alpha\alpha',\beta\beta'}
&=&
d^{\bf S}_{\alpha\beta}
\cdot\left( \mbox{\boldmath$\cal B$}^{\rm S} \overrightarrow{\nabla}^{\rm S}
H_{\rm S}\right) \delta_{\beta'\alpha'}
+
\frac{1}{2}\Delta E_{\alpha\beta}
d^{\bf S}_{\alpha\beta}
\cdot\left(
\mbox{\boldmath$\cal B$}^{\rm S}
\overrightarrow{\nabla}^{\rm S}
\right)
\delta_{\alpha'\beta'}
\\
&+& 
d^{{\bf S}*}_{\alpha'\beta'} 
\cdot
\left( \mbox{\boldmath$\cal B$}^{\rm S}
\overrightarrow{\nabla}^{\rm S} H_{\rm S} \right)
\delta_{\alpha\beta}
+\frac{1}{2}\Delta E_{\alpha'\beta'}
d^{{\bf S}*}_{\alpha'\beta'}
\cdot
\left( \mbox{\boldmath$\cal B$}^{\rm S} \overrightarrow{\nabla}^{\rm S} \right)
\delta_{\alpha\beta}
\;.
\end{array}
\label{eq:Jspin}
\end{eqnarray}

The limit $d_{\alpha\alpha}^{\bf S}\to 0$ of the spin transition operator
in Equation~(\ref{eq:Jspin})
provides the form of the standard transition operator
for canonical conjugate coordinates,
given in Equation~(\ref{eq:T_aa'bb'}).
Finally, because of the spin nature of the bath, one finds a higher order 
transition operator (which does not appear in the case of canonical conjugate 
bath coordinates):
\begin{eqnarray}
{\cal S}_{\alpha\alpha',\beta\beta'}
&=&
\frac{1}{2} \Delta E_{\alpha\sigma} d^{\bf S}_{\alpha\sigma}
\mbox{\boldmath$\cal B$}^{\rm S} d^{\bf S}_{\sigma\beta}\delta_{\alpha'\beta'}
+\frac{1}{2} \Delta E_{\alpha\beta} d^{\bf S}_{\alpha\beta}
\mbox{\boldmath$\cal B$}^{\rm S} d^{{\bf S}*}_{\alpha'\beta'}
\nonumber\\
&+& \frac{1}{2}\Delta E_{\alpha'\sigma'} d^{{\bf S}*}_{\alpha'\sigma'}
\mbox{\boldmath$\cal B$}^{\rm S}
d^{{\bf S}*}_{\sigma'\beta'}\delta_{\alpha\beta}
+\frac{1}{2}\Delta E_{\alpha'\beta'} d^{{\bf S}*}_{\alpha'\beta'}
\mbox{\boldmath$\cal B$}^{\rm S} d^{\bf S}_{\alpha\beta}
\nonumber\\
&-&\frac{1}{2}\left(E_{\alpha}+E_{\alpha'}\right)\overleftarrow{\nabla}^{\bf S}
\mbox{\boldmath$\cal B$}^{\rm S} \cdot
d^{\bf S}_{\alpha\beta}\delta_{\alpha'\beta'}
-\frac{1}{2} \left(E_{\alpha}+E_{\alpha'}\right)
\overleftarrow{\nabla}^{\bf S} \mbox{\boldmath$\cal B$}^{\rm S}
\cdot d^{{\rm S}*}_{\alpha'\beta'}\delta_{\alpha\beta}
\;.
\label{eq:Sspin}
\end{eqnarray}

The adiabatic limit of the spin-bath QCLE in~(\ref{eq:qcsd-dyna}) 
can be taken by setting to zero
the off-diagonal elements of $d_{\alpha\alpha'}$,
which appear in the operators in Equations~(\ref{eq:Jspin}) and~(\ref{eq:Sspin}).
This is physically reasonable whenever the coupling
between the different adiabatic energy surfaces is negligible.
One obtains
\begin{eqnarray}\begin{array}{lll}
{\cal T}^{{\bf S},{\rm ad}}_{\alpha\alpha',\beta\beta'}
&=&
\left( d^{\bf S}_{\alpha\alpha} + d^{{\bf S}*}_{\alpha'\alpha'} \right)
\mbox{\boldmath$\cal B$}^{\bf S} \overrightarrow{\nabla}^{\bf S} H_{\rm S}
\delta_{\alpha\beta}\delta_{\beta'\alpha'}
\\
&=&
- i
\left( \phi^{\bf S}_{\alpha\alpha} - \phi^{\bf S}_{\alpha'\alpha'} \right)
\mbox{\boldmath$\cal B$}^{\bf S} \overrightarrow{\nabla}^{\bf S}
\delta_{\alpha\beta}\delta_{\beta'\alpha'}
\;. \end{array}
\label{eq:Jspin-ad}
\end{eqnarray}

The geometric phase
\begin{equation}
\phi^{\rm S}_{\alpha\alpha}=-id^{\bf S}_{\alpha\alpha}
\label{eq:geo-phase}
\end{equation}
has been introduced exploiting the
purely imaginary character of $d^{\bf S}_{\alpha\alpha}$.
Similarly, the higher order transition operator becomes
\begin{eqnarray}
{\cal S}_{\alpha\alpha',\beta\beta'}^{\rm ad}
&=&
-\frac{i}{2}
\sum_{I,J}
\left(\phi^{\bf S}_{\alpha\alpha} - \phi^{\bf S}_{\alpha'\alpha'}\right)
\mbox{\boldmath$\cal B$}^{\rm S} \overrightarrow{\nabla}^{\bf S}
\left(E_{\alpha}+E_{\alpha'}\right)\delta_{\alpha\alpha}\delta_{\alpha'\alpha'}
\nonumber\\
\label{eq:Sspin-ad}
\end{eqnarray}

Putting everything together, the adiabatic approximation of the spin-bath QCLE may be written~as
\begin{eqnarray}
\frac{\partial}{\partial t}W^{\bf S}_{\alpha\alpha'} ({\bf S},t)
&=&\left[-i\omega_{\alpha\alpha'}
-i 
\left(\phi^{\bf S}_{\alpha\alpha} - \phi^{\bf S}_{\alpha'\alpha'}\right)
\mbox{\boldmath$\cal B$}
\overrightarrow{\nabla}^{\bf S}
H_{\bf S}^{\alpha\alpha'}
- H_{\rm S}^{\alpha\alpha'}
\overleftarrow{\nabla}^{\bf S}
\mbox{\boldmath$\cal B$}
\overrightarrow{\nabla}^{\bf S}
\right]
W^{\bf S}_{\alpha\alpha'}({\bf S},t).
\label{eq:qcsd-dyna-ad}
\end{eqnarray}

In Equation~(\ref{eq:qcsd-dyna-ad}), 
the phase $\omega_{\alpha\alpha'}$ has a dynamical nature
while the phase $\phi_{\alpha\alpha}^{\bf S}$ is of a geometric origin
and it can be considered an instance of
the famous Berry phase~\cite{berry,qphases,mead}.
Interestingly, Equation~(\ref{eq:pWS}) predicts
that the geometric phase $\phi_{\alpha\alpha}^{\bf S}$
can be non-zero also for open paths of the classical spins of the bath 
(open-path Berry phases were discussed in Ref.~\cite{pati}).
Moreover, the phase factor $\phi_{\alpha\alpha}^{\bf S}-\phi_{\alpha'\alpha'}^{\bf S}$
is~purely off-diagonal (off-diagonal Berry phases
for environments described by canonically conjugate
variables were discussed in Refs.~\cite{filipp,englman,manini}).
It is worth mentioning that the geometric phase $\phi_{\alpha\alpha}^{\bf S}$
is predicted also for non-adiabatic dynamics.

When the total Hamiltonian is time-independent, as the one
in Equation~(\ref{eq:tot-hamS}), the adiabatic evolution of the
matrix elements of the spin-bath operator-valued
quasi-probability function, given by Equation~(\ref{eq:qcsd-dyna-ad}),
can be rewritten as
\begin{eqnarray}
\frac{\partial}{\partial t}W^{\bf S}_{\alpha\alpha'}({\bf S},t)
&=&\left[-i\omega_{\alpha\alpha'}
-\left( \langle\alpha,S|\frac{d}{dt}|\alpha,S\rangle -\langle\alpha',S|\frac{d}{dt}|\alpha',S\rangle \right)
- H^{\alpha\alpha'}_{\rm S}
\overleftarrow{\nabla}^{\bf S}
\mbox{\boldmath$\cal B$}^{\bf S}
\overrightarrow{\nabla}^{\bf S}
\right]
W^{\bf S}_{\alpha\alpha'}({\bf S},t) \;. \nonumber\\
\label{eq:qcsd-dyna-ad2}
\end{eqnarray}

Using the Dyson identity, one can obtain the following form for 
$\hat W^{\bf S}({\bf S},t)$
in terms of the adiabatic propagator:
\begin{eqnarray}
W^{\bf S}_{\alpha\alpha'}({\bf S},t)
&=&\exp\left[-i\int_{t_0}^t dt'\omega_{\alpha\alpha'}(t')\right]
\exp\left[
-\int_{t_0}^t dt'\left( \langle\alpha,S|\frac{d}{dt'}|\alpha,S\rangle
 -\langle\alpha',S|\frac{d}{dt'}|\alpha',S\rangle \right)\right]
\nonumber\\
&\times&
\exp\left[ -(t-t_0)
H_{\alpha\alpha'}^{\rm S}
\overleftarrow{\nabla}^{\bf S}
\mbox{\boldmath$\cal B$}^{\rm S}
\overrightarrow{\nabla}^{\bf S}
\right]
W^{\bf S}_{\alpha\alpha'}({\bf S},t_0)\;.
\label{eq:qcsd-dyna-ad-prop}
\end{eqnarray}
Equation~(\ref{eq:qcsd-dyna-ad-prop}) 
provides a convenient starting point for devising numerical
integration schemes based on the SSTP propagation scheme~\cite{sstp}.

In Ref.~\cite{sergi-spin2}, the following model Hamiltonian was considered:
\begin{eqnarray}
\hat H({\bf S}) &=& -\Omega \hat\sigma_x-c_1b\hat\sigma_z
-\mu{\bf S}\cdot\mbox{\boldmath$\sigma$}-c_2bS_z+\frac{S_z^2}{2}
\\
&=&\hat h_{\bf S}({\bf S}) -c_2bS_z +\frac{S_z^2}{2}\;,
\end{eqnarray}
where $\Omega$, $c_1$, and $c_2$ are real parameters,
$b$ is the $z$ component of the magnetic field ${\bf B}=(0,0,b)$,
while $\mbox{\boldmath$\sigma$}=(\sigma_x,\sigma_y,\sigma_z)$
is a vector having the Pauli matrices $\sigma_x$,
$\sigma_y$, and $\sigma_z$ as components.
The SSTP algorithm was applied to Equation~(\ref{eq:qcsd-dyna-ad-prop})
and the action of the classical like Liouville operator
$H_{\alpha\alpha'}^{\rm S} \overleftarrow{\nabla}^{\bf S}
\mbox{\boldmath$\cal B$}^{\rm S} \overrightarrow{\nabla}^{\bf S}$
was evaluated using time reversible integration algorithms
based on the symmetric break-up of
the Liouville propagator~\cite{tuck1,tuck2,ribes}.


\section{Stochastic Classical Baths}
\label{sec:qcd-sb}

Consider a quantum-classical system comprising a quantum subsystem
and a classical environment whose classical phase
space coordinates are partitioned into two sets: one set
$X=(Q,P)$ interacts directly with the quantum subsystem
while the second set $X'=(Q^{\prime},P^{\prime})$
interacts only with the coordinates $X$
(and therefore is not directly coupled to the quantum subsystem).
We assume that the detailed dynamics of the coordinates $X'$
is not interesting: their function is just that of working as a thermal bath,
leading to dissipative dynamics~\cite{ray-barbados}.

An equation of motion for the hybrid quantum-classical system composed of
the quantum subsystem and the classical DOF $X$
only has been derived using projection operator
methods~\cite{ray-barbados}. It~takes the form,
\begin{eqnarray}
\frac{\partial}{\partial t}
\hat W(X,t)
&=&
-\frac{i}{\hbar}
\left[\begin{array}{cc} \hat H _{\rm W} & \hat W(X,t)\end{array}\right]
\mbox{\boldmath$\cal D$}
\left[\begin{array}{c} \hat H _{\rm W} \\ \hat W(X,t)\end{array}\right]
\nonumber \\
& +&\zeta\overrightarrow{\nabla}_P\left(\frac{P}{M}
+k_BT\overrightarrow{\nabla}_P\right) \hat W(X,t)\;,
=-i\hat{\cal L}^{\rm D}\hat W(X,t) \;, \label{stocha-qc-liouville}
\end{eqnarray}
where $\nabla_P=\partial/\partial P$, $\zeta$ is the friction constant,
$k_B$ is the Boltzmann
constant, and $T$ is the temperature of the bath.
The Hamiltonian in Equation~(\ref{stocha-qc-liouville})
is defined in Equation~(\ref{eq:H_W}).
However, in the present case, we~must interpret $V_{\rm W}(\hat q,Q)$
as the potential of mean force arising
from the average over the primed bath variables $Q'$.
The Liouville operator $i\hat{\cal L}^{\rm D}$,
defined on the right-hand side of Equation~(\ref{stocha-qc-liouville}),
determines the dissipative dynamics of the system.
This Fokker--Planck-like operator and the potential of mean force make
the dissipative quantum-classical Liouville operator in
Equation~(\ref{stocha-qc-liouville}) different from that 
describing an isolated quantum-classical system~\cite{kapra-cicco}.
In particular, the term $\zeta\overrightarrow{\nabla}_P\left[(P/M)
+k_BT\overrightarrow{\nabla}_P\right]$ directly breaks the time-translation
symmetry leading to diffusive motion and energy dissipation.

The dissipative Liouville operator can be written in the adiabatic basis as
\begin{equation}
i\hat{\cal L}^{\rm D}_{\alpha\alpha'\beta\beta'}=
\left( i\omega_{\alpha\alpha'}(R)
+iL_{\alpha\alpha'}^{\rm K}\right)
\delta_{\alpha\beta}\delta_{\alpha'\beta'}
+{\cal T}_{\alpha\alpha'\beta\beta'}
\;,
\end{equation}
where we have defined the Kramers operator as
\begin{eqnarray}
i L_{\alpha\alpha'}^{\rm K}&=&
\left[ \frac{P}{M}\overrightarrow{\nabla}_Q +\frac{1}{2}\left(F_W^{\alpha}
+F_W^{\alpha'}\right) \overrightarrow{\nabla}_P
-\zeta\overrightarrow{\nabla}_P
\left(\frac{P}{M}+k_BT\overrightarrow{\nabla}_P \right) \right]
\;.
\end{eqnarray}

The quantum-classical average of any operator or dynamical
variable $\hat \chi(X)$ can be written as
\begin{eqnarray}
\langle\hat \chi\rangle(t)
&=&\sum_{\alpha\alpha'\beta\beta'}\int dX
\chi_{\alpha'\alpha}(X)
\exp[-i{\cal L}_{\alpha\alpha'\beta\beta'}^Dt]
W^{\beta\beta'}(X)
\nonumber \\
&=&\sum_{\alpha\alpha'\beta\beta'}\int dX W^{\beta\beta'}(X)
\exp[i{\cal L}_{\beta'\beta\alpha'\alpha}^{\rm DB}t]
\chi_{\alpha'\alpha}(R,P),
\end{eqnarray}
where $i{\cal L}_{\beta'\beta\alpha'\alpha}^{\rm DB}$ is the backward
operator, defined as
\begin{eqnarray}
i\hat{\cal L}^{DB}_{\alpha\alpha'\beta\beta'}&=&
\left(i\omega_{\alpha\alpha'}(R)
+iL_{\alpha\alpha'}^{\rm KB}\right)
\delta_{\alpha\beta}\delta_{\alpha'\beta'}
+{\cal T}_{\alpha\alpha'\beta\beta'}
\;
\end{eqnarray}

The backward Kramers $iL_{\alpha\alpha'}^{\rm KB}$ operator
is written as
\begin{eqnarray}
i L_{\alpha\alpha}^{\rm KB}&=&
\left[
\frac{P}{M}\overrightarrow{\nabla}_Q
+\frac{1}{2}\left(F_{\rm W}^{\alpha}+F_{\rm W}^{\alpha'}\right)
\overrightarrow{\nabla}_P
-\zeta \left(\frac{P}{M}-k_BT\overrightarrow{\nabla}_P\right)
\overrightarrow{\nabla}_P\right]
\delta_{\alpha\beta}\delta_{\alpha'\beta'}
\;.
\end{eqnarray}

According to the classical theory of random processes~\cite{gardiner},
the time evolution under the backward Kramers operator
$i{\cal L}_{\alpha\alpha'\beta\beta'}^{KB}$ can be unfolded {it via} an
average over realizations of stochastic Langevin trajectories.
In such a  picture, the classical trajectory segments obey 
the Langevin equations of motion,
\begin{eqnarray}
\dot Q&=& \frac{P}{M},\label{eq:r-lang-1}\\
\dot P &=&-\frac{\zeta}{M}P +\frac{1}{2}
\left(F_{\rm W}W^{\alpha}+F_{\rm W}^{\alpha'}\right)
+ {\cal R}(t)\;,\label{eq:p-lang-2}
\end{eqnarray}
where ${\cal R}(t)$ is a Gaussian white noise process with the
properties,
\begin{eqnarray}
\langle {\cal R}(t)\rangle &=& 0\;,\\
\langle {\cal R}(t){\cal R}(t')\rangle &=& 2k_BT\zeta\delta(t-t')
\;.
\end{eqnarray}

To Equations~(\ref{eq:r-lang-1}) and (\ref{eq:p-lang-2}),
one can associate a time-dependent Langevin--Liouville operator
\begin{equation}
iL_{\alpha\alpha'}^{\rm L}(t)=
\frac{P}{M}\overrightarrow{\nabla}_Q
+\left(-\frac{\zeta}{M}P
+\frac{1}{2}(F_{\rm W}^{\alpha}+F_{\rm W}^{\alpha})
+ {\cal R}(t)\right) \overrightarrow{\nabla}_P \;,
\label{L_L-op}
\end{equation}
and a time-ordered propagator
\begin{equation}
U^{\rm L}_{\alpha\alpha'}(t,0)={\cal T}\exp\left[\int_0^tdt'
iL^{\rm L}_{\alpha\alpha'}(t')\right]
\;. \label{eq:lang-prop}
\end{equation}

In order to generate the stochastic Langevin trajectories, we can use
a total time-dependent Langevin--Liouville super-operator
\begin{equation}
i\hat{\cal L}_{\alpha\alpha'\beta\beta'}^{\rm L}(t)=
\left(i\omega_{\alpha\alpha'}(Q) +i L_{\alpha\alpha'}^{\rm L}(t)\right)
\delta_{\alpha\beta}\delta_{\alpha'\beta'}
+{\cal T}_{\alpha\alpha'\beta\beta'}
\label{ll-op}
\end{equation}
and the associated propagator
\begin{equation}
{\cal U}_{\alpha\alpha'\beta\beta'}^{\rm L}(t,0)=
{\cal T}\exp\left[\int_0^tdt'
i{\cal L}_{\alpha\alpha'\beta\beta'}^{\rm L}(t')\right]
\;.
\label{ll-prop}
\end{equation}
Within such a Langevin picture, the quantum-classical average of
any operator $\hat\chi(X)$ can be calculated~as \vspace{12pt}
\begin{eqnarray}
\langle\hat{\chi}\rangle(t)
&=&\sum_{\alpha\alpha'\beta\beta'}\int dX
W^{\beta\beta'}(Q)
\overline{{\cal U}_{\beta\beta'\alpha\alpha'}^{\rm L}(t)
\chi_{\alpha'\alpha}(Q)
} \; \end{eqnarray}
where the over-line denotes an average over an ensemble of
stochastic Langevin trajectories.
Since~they are independent from each other,
the order in which the average
over phase space and the average over the stochastic Langevin process
are performed can be permuted.
Hence, one can write
\begin{eqnarray}
\langle\hat\chi(X,t)\rangle
&=&\sum_{\alpha\alpha'\beta\beta'}
\overline{
\int dRdP W^{\beta\beta'}(X)
{\cal U}_{\beta\beta'\alpha\alpha'}^{\rm L}(t)
\chi_{\alpha'\alpha}^{\prime}(XP)
}.\nonumber \\
\label{stocha-av}
\end{eqnarray}
Equation~(\ref{stocha-av}) allows one to calculate
averages in a quantum-classical dissipative system
as phase space weighted averages over many 
Langevin trajectories.

In Ref.~\cite{langevin-qc}, a quantum subsystem with two energy levels
interacting with a dissipative classical quartic oscillator was considered.
The Hamiltonian of the hybrid quantum-classical system reads
\begin{eqnarray}
\hat{H}_W(X)&=&\frac{P^2}{2M}+V_q(Q)-\hbar\Omega\hat{\sigma}_x
-\hbar\gamma_0Q\hat{\sigma}_z
\label{eq:ham-model}
\;,
\end{eqnarray}
where $V_q(Q)=\frac{a}{4}R^4-\frac{b}{2}R^2$,
$\Omega$, $a$, $b$, and $\gamma_0$ are real parameters, $M$ is the
mass of the quartic oscillator, and $\hat{\sigma}_x$
and $\hat{\sigma}_z$ are Pauli matrices.

The calculation of quantum-classical averages
using the dynamics defined by the time-dependent
Langevin--Liouville propagator ${\cal U}_{ss'}^L(t)$ in Equation~(\ref{ll-prop})
is no more complicated than that for deterministic quantum-classical
dynamics. The momentum-jump approximation~\cite{simu,sergi-theor} and
a simple generalization of the SSTP algorithm~\cite{sstp,sergi-theor}
to the time dependent propagator were used in Ref.~\cite{langevin-qc}.
The explicitly time-dependent propagator ${\cal U}_{ss'}^L(t)$ must be defined
as a time ordered product.
A simple way to achieve that is to employ the decomposition scheme devised by
Suzuki~\cite{suzuki}.
Details of the numerical procedures are found in Ref.~\cite{langevin-qc}


\section{Non-Hamiltonian Dynamics in Thermal Baths}
\label{sec:qcnose}

By exploiting the antisymmetric structure of the quantum-classical commutator,
arising from the matrix operator $\mbox{\boldmath$\cal D$}$
given in Equation~(\ref{eq:D_matrix}), one can
impose the thermodynamic constraints of constant temperature
on the classical-like DOF~\cite{b3,b4}.
Following Refs.~\cite{b1,b2,aspvg},
constant-temperature dynamics for the classical bath coordinates,
as defined through the non-Hamiltonian Nos\'e--Hoover
equations of motion, can be introduced by modifying the
matrix $\mbox{\boldmath$\cal B$}$ and augmenting in a minimal
way the dimension of the phase space bath.
The classical Nos\'e--Hoover thermostat is briefly discussed in
Appendix~\ref{app:nh}.

As in the classical case, the Nos\'e variables are
\begin{equation}
X^{\rm N}\equiv (Q,Q_\eta,P,P_{\eta})\;,
\end{equation}
where $Q_\eta$ and $P_{\eta}$ are the Nos\'e coordinate and momentum.
The Nos\'e quantum-classical Hamiltonian is obtained
by adding the Nos\'e kinetic energy $P_{\eta}^2/2M_\eta$
and potential energy $Nk_BTQ_\eta$ to $\hat H_{\rm W}$
in Equation~(\ref{eq:H_W})
\begin{equation}
H^{\rm N}=
\frac{P^2}{2M}+\frac{P_{\eta}^2}{2M_{\eta}} + Nk_{\rm B}TQ_\eta
+\hat h_{\rm W}(Q) \;,
\end{equation}
where $M_{\eta}$ is the Nos\'e inertial parameter, $k_{\rm B}$
is the Boltzmann constant, $T$ is the constant temperature,
and~$N$ is the number of $Q$ coordinates.
Using the matrix $\mbox{\boldmath$\cal B$}^{\rm N}$ 
in Equation~(\ref{eq:BN_matrix}),
the classical phase space quasi-Hamiltonian bracket
of two variables $A_1$ and $A_2$ can be defined as
\begin{equation} 
A_1 \overleftarrow{\nabla}^{\rm N}
\mbox{\boldmath$\cal B$}^{\rm N}
\overrightarrow{\nabla}^{\rm N} A_2
=\sum_{I,J=1}^{2(N+1)}
A_1 \overleftarrow{\nabla}_I^{\rm N}
{\cal B}_{IJ}^{\rm N} \overrightarrow{\nabla}_J^{\rm N} A_2 \;.
\label{gen-bracket}
\end{equation}

The explicit form of the matrix operator,
which defines the quantum-classical bracket
and the law of motion  through Equation~(\ref{eq:qcle}),
is then given by
\begin{equation}
\mbox{\boldmath$\cal D$}^{\rm N}=
\left[
\begin{array}{cc} 0 & 1
-\frac{\overleftarrow{\nabla}^{\rm N}{\cal B}^{\rm N}
\overrightarrow{\nabla}^{\rm N}}
{2i\hbar^{-1}}\\
-\left(1
-\frac{\overleftarrow{\nabla}^{\rm N}{\cal B}^{\rm N}
\overrightarrow{\nabla}^{\rm N}}
{2i\hbar^{-1}}\right) & 0
\end{array}
\right]\;.
\end{equation}

The Nos\'e--Hoover QCLE for the operator-valued quasi-probability
function $\hat W^{\rm N}(X^{\rm N},t)$ is given~by
\begin{eqnarray}
\frac{d}{dt}\hat W^{\rm N}(X^{\rm N},t)
&=&
-i{\cal L}^{\rm N} W^{\rm N}(X^{\rm N},t)
-\kappa^{\rm N}(X^{\rm N})W^{\rm N}(X^{\rm N},t)
\nonumber\\
&=&
-\frac{i}{\hbar}\left[\begin{array}{cc}\hat H^{\rm N} &
\hat W^{\rm N}(X^{\rm N},t)
 \end{array}\right]
\cdot\mbox{\boldmath$\cal D$}^{\rm N}\cdot
\left[\begin{array}{c} \hat H^{\rm N}\\ \hat W^{\rm N}(X^{\rm N},t)
\end{array}\right]
-\kappa^{\rm N}(X^{\rm N})W^{\rm N}(X^{\rm N},t) \;.
\label{eq:nose-eqofm}
\end{eqnarray}

The presence of the term $-\kappa^{\rm N}(X^{\rm N})W^{\rm N}(X^{\rm N},t)$
in the left-hand side of Equation~(\ref{eq:nose-eqofm})
derives from the passage from the Heisenberg to the
Schr\"odinger picture, as it is explained in Appendix~\ref{app:stationary}.

Upon considering the term 
in the right-hand side of (\ref{eq:nose-eqofm}), 
one obtains
\begin{eqnarray}
\hat H^{\rm N}
\overleftarrow{\nabla}^{\rm N}
{\cal B}^{\rm N} \overrightarrow{\nabla}^{\rm N}
\hat{\chi}(X^{\rm N},t)
-\hat{\chi}(X^{\rm N},t)
\overleftarrow{\nabla}^{\rm N}{\cal B}^{\rm N}\overrightarrow{\nabla}^{\rm N}
\hat{H}_{\rm N}
&=&
\frac{\partial\hat V}{\partial Q}
\frac{\partial\hat\chi(X^{\rm N},t)}{\partial P}
+\frac{\partial\hat\chi(X^{\rm N},t)}{\partial P}
\frac{\partial\hat V}{\partial Q}
\nonumber\\
&-&2F_{Q_\eta}\frac{\partial\hat\chi(X^{\rm N},t)}{\partial P_{\eta}}
-2\frac{P}{M}\frac{\partial\hat\chi(X^{\rm N},t)}{\partial Q}
 \\
&-&2\frac{P_{\eta}}{M_{\eta}}
\frac{\partial\hat\chi(X^{\rm N},t)}{\partial Q_\eta}
+2\frac{P_{\eta}}{M_{\eta}}P
\frac{\partial\hat\chi(X^{\rm N},t)}{\partial P}
\;,\nonumber
\end{eqnarray}
where $F_{Q_\eta}=\frac{P^2}{M}-Nk_BT$.
Finally, using the above result,
the Nos\'e--Hoover QCLE reads
\begin{eqnarray}\begin{array}{lll}
 \small
\frac{d}{dt}\hat W^{\rm N}(X^{\rm N},t)
&=&
-\frac{i}{\hbar}\left(H^{\rm N}\hat W^{\rm N}(X^{\rm N},t)
-\hat\chi(X^{\rm N},t)H^{\rm N}\right)
+\frac{1}{2}
\left(\frac{\partial\hat W^{\rm N}(X^{\rm N},t)}{\partial P}
\frac{\partial\hat V}{\partial Q}
+\frac{\partial\hat V}{\partial Q}
\frac{\partial\hat\chi(X^{\rm N},t)}{\partial P}\right)
 \\
&&
-\frac{P}{M}\frac{\partial\hat W^{\rm N}(X^{\rm N},t)}{\partial Q}
-\frac{P_{\eta}}{M_{\eta}}
\frac{\partial\hat\chi(X^{\rm N},t)}{\partial Q_\eta}
+\frac{P_{\eta}}{M_{\eta}}P
\frac{\partial\hat\chi(X^{\rm N},t)}{\partial P}
-F_{Q_\eta}\frac{\partial\hat W^{\rm N}(X^{\rm N},t)}{\partial P_{\eta}}
\;.
 \end{array}
\label{eq:noseeqofm}
\end{eqnarray}

In the adiabatic states defined in Equation~(\ref{eq:adia_states}),
Equation~(\ref{eq:noseeqofm}) reads
\begin{equation}
\frac{d}{dt}\hat W^{\rm N}_{\alpha\alpha'}(X^{\rm N},t) =
-\sum_{\beta\beta'} i{\cal L}^{\rm N}_{\alpha\alpha',\beta\beta'}
\hat W^{\rm N}_{\beta\beta'}(X^{\rm N},t)
\;,
\end{equation}
where
\begin{equation}
i{\cal L}^{\rm N}_{\alpha\alpha',\beta\beta'}
=
i\omega_{\alpha\alpha'}\delta_{\alpha\beta}\delta_{\alpha'\beta'}
+
\delta_{\alpha\beta}\delta_{\alpha'\beta'}i{L}^{\rm N}_{\alpha \alpha '}
+
{\cal T}_{\alpha \alpha ',\beta\beta'}\;.
\label{eq:nose-liouville}
\end{equation}

We have used the definition of the Bohr frequency $\omega_{\alpha\alpha'}$
in Equation~(\ref{eq:Bohr-omega}) and of the transition operator
${\cal T}_{\alpha\alpha',\beta\beta'}$ in Equation~(\ref{eq:T_aa'bb'})
in Appendix~\ref{sec:adbas_rep}.
We have introduced
a classical-like Nos\'e--Liouville~operator
\begin{eqnarray}
i\hat{L}^{\rm N}_{\alpha\alpha '}&=&
\frac{P}{M}\frac{\partial }{\partial Q}
+\frac{1}{2}\left(F^{\alpha}+F^{\alpha'}\right)
\nonumber\\
&-&P\frac{P_{\eta}}{M_{\eta}}\frac{\partial}{\partial P}
+\frac{P_{\eta}}{M_{\eta}}\frac{\partial}{\partial Q_\eta}
+F_{Q_\eta}\frac{\partial}{\partial P_{\eta}} \frac{\partial}{\partial P}
\;.
\end{eqnarray}

The existence of the stationary operator-valued Nos\'e quasi-probability
function $\hat W^{\rm N,e}(X^{\rm N})$
is discussed in Appendix~\ref{app:stationary}.

\subsection{Nos\'e--Hoover Chain Thermal Baths}

The Nos\'e--Hoover thermostat suffers from lack of ergodic dynamics
when the bath has high frequencies of motion.
The Nos\'e--Hoover chain~\cite{nhc} is a more general non-Hamiltonian
thermostat that solves the ergodicity problems suffered by
the standard Nos\'e--Hoover thermostat in the case of stiff variables.
The Nos\'e--Hoover chain thermostat can also be formulated in a
quantum-classical framework with minimal changes with respect to
what is shown in Section~\ref{sec:qcnose}.
To this end, considering for simplicity a chain of just two
thermostat coordinates,
one can define the classical phase space point~as
\begin{equation}
X^{\rm NHC}=(R,Q_{\eta_1},Q_{\eta_2},P,P_{\eta_1},P_{\eta_2})
\;,
\end{equation}
\begin{eqnarray}
\hat{H}^{\rm NHC}&=&\frac{\hat{p}^2 }{2m}+\frac{P^2 }{2M}
+\frac{P_{\eta_1}^2}{2M_{\eta_1}}
+\frac{P_{\eta_2}^2}{2M_{\eta_2}}
\nonumber \\
&+&\hat V(\hat q,R) + Nk_BTQ_{\eta_1} +k_BTQ_{\eta_2}
\;,
\end{eqnarray}
where $M_{\eta_1}$ and $M_{\eta_2}$ are the inertial
parameters of the thermostat variables.
As shown in  Ref.~\cite{b1,b2},
one can define an antisymmetric matrix
\begin{equation}
\mbox{\boldmath$\cal B$}^{\rm NHC}=
\left[
\begin{array}{cccccc}
0 & 0 & 0 & 1 & 0 & 0 \\
0 & 0 & 0 & 0 & 1 & 0 \\
0 & 0 & 0 & 0 & 0 & 1 \\
-1 & 0 & 0 & 0 & -P & 0 \\
0 & -1 & 0 & P& 0 & -P_{\eta_1}\\
0 & 0 & -1 & 0 & P_{\eta_1} & 0
\end{array}
\right]\;.
\end{equation}

The matrix $\mbox{\boldmath$\cal B$}^{\rm NHC}$ 
can be used to define the quasi-Hamiltonian bracket
according to Equation~(\ref{eq:symplectic-bracket}).
The Nos\'e--Hoover chain classical equations of motion
in phase space~\cite{b1} are then given by
\begin{equation}
\dot{X}=-X^{\rm NHC}
\overleftarrow{\nabla}^{\rm NHC}{\cal B}^{\rm NHC}
\overrightarrow{\nabla}^{\rm NHC}\hat H^{\rm NHC}.
\end{equation}

Quantum-classical dynamics is then introduced
using the matrix super-operator
\begin{equation}
\mbox{\boldmath$\cal D$}^{\rm NHC}=\left[\begin{array}{cc}
0 & 1-
\frac{\overleftarrow{\nabla}^{\rm NHC}{\cal B}^{\rm NHC}
\overrightarrow{\nabla}^{\rm NHC}}{2i\hbar^{-1}}
\\
-\left(1-
\frac{\overleftarrow{\nabla}^{\rm NHC}{\cal B}^{\rm NHC}
\overrightarrow{\nabla}^{\rm NHC}}{2i\hbar^{-1}}
\right) & 0\end{array}\right].
\end{equation}

The quantum-classical equations of motion
can then be written as
\begin{eqnarray}
\frac{d\hat{\chi}}{dt}&=&\frac{i}{\hbar}
\left[\begin{array}{cc}\hat{H}^{\rm NHC} &\hat{\chi}\end{array}\right]
\cdot\mbox{\boldmath$\cal D$}^{\rm NHC}\cdot
\left[\begin{array}{c}\hat{H}^{\rm NHC} \\ \hat{\chi}\end{array}\right].
\end{eqnarray}

The equations of motion can be represented using the
adiabatic basis obtaining the Liouville super-operator
\begin{eqnarray}
i{\cal L}_{\alpha\alpha',\beta\beta'}^{\rm NHC}&=&
(i\omega_{\alpha\alpha'}+iL_{\alpha\alpha'}^{\rm NHC})\delta_{\alpha\beta}
\delta_{\alpha'\beta'} -{\cal T}_{\alpha\alpha',\beta\beta'}
\;, \nonumber \\
\end{eqnarray}
where
\begin{eqnarray}
iL_{\alpha\alpha'}^{\rm NHC}&=&\frac{P}{M}\frac{\partial}{\partial R}
+\frac{1}{2}(F^{\alpha}+F^{\alpha'})\frac{\partial}{\partial P}
+\sum_{k=1}^2(\frac{P_{\eta_k}}{M_{\eta_k}}
\frac{\partial}{\partial Q_{\eta_k}}
+F_{Q_{\eta_k}}\frac{\partial}{\partial P_{\eta_k}})
- \frac{P_{\eta_2}}{M_{\eta_2}}
P_{\eta_1}\frac{\partial}{\partial P_{\eta_1}}
\;,
\end{eqnarray}
with $F_{Q_{\eta_2}}=(P_{\eta_1}^2/M_{\eta_1})-k_BT$.
The proof of the existence of stationary density matrix 
in the case of Nos\'e--Hoover chains follows
the same logic of the simpler Nos\'e--Hoover case.
In the adiabatic basis,
the density matrix stationary up to order bar
has the same form as that given in Equations~(\ref{ansatz})
and~(\ref{ansatz2}). One has just to replace
Equation~(\ref{ansatz}) for the order zero term with
\begin{eqnarray}
W{\alpha\alpha}^{{\rm NHC,e},(0)}
&=&\frac{1}{Z^{\rm NHC}}
e^{-\beta\left[\frac{P^2}{2M}+E_{\alpha}(R)
+\sum_{k=1}^2\left(\frac{P_{\eta_k}^2}{2M_{\eta_k}}
\right) +Nk_BTQ_{\eta_1} +k_BTQ_{\eta_2}
\right]}
\end{eqnarray}
with an obvious definition of $Z^{\rm NHC}$.


\section{Conclusions and Perspectives}
\label{sec:cp}

In this review, we discussed how to mathematically describe the dynamics
and statistical mechanics of quantum subsystems embedded in classical baths.
The formalism is founded on an operator-valued quasi-probability function
evolving through a QCLE defined in terms of a quasi-Lie bracket.
 {It is worth emphasizing that the QCLE is a fully
deterministic equation that takes into account explicitly
{\it all} the DOF of the system, {i.e.}, it
describes the quantum and classical DOF of the total hybrid system.
Hence, the QCLE generates a unitary dynamics, conserving both the
system's probability and energy.}
However, the time-translation invariance of the quasi-Lie bracket algebra is broken.
This situation is surprising:
one does not expect a broken time-translation invariance symmetry
in an isolated system when all its degrees of freedom are
taken into account. This can be seen as a signature of the effect
of the classical bath on the quantum subsystem, and of the back-reaction
of the subsystem onto the bath. In other words, the total hybrid system is closed
from the point of view of energy and probability conservation but,
because of the above mentioned back-reaction,
it is also open: the quasi-Lie bracket
describes the irreversible transfer of quantum
information onto the classical DOF.
We also reviewed how the hybrid quantum-classical theory
can be derived from a partial Wigner transform and a semiclassical
limit of the QLE only in the case when the bath is described by
canonically conjugate coordinates.
After this, we discussed how to treat quantum subsystems 
embedded in both non-canonical and non-Hamiltonian bath.
In all cases, the mathematical object representing the state of the
system is an operator-valued quasi-probability function
that depend on the coordinates of the bath and
whose equation of motion depends on the specific case under study.
It~is explained how classical spin baths
are described in terms of non-canonical coordinates and how this fact
leads to the appearance of an off-diagonal open-path
geometric phase in the dynamics of the operator-valued quasi-probability
function of the system.
We then discussed how the effect of
thermal baths can be implemented by means of a stochastic, quantum-classical
Langevin dynamics and by means of a
deterministic, non-Hamiltonian Nos\'e--Hoover thermostatted dynamics.
The formulation of the dynamics in both the spin
and Nos\'e--Hoover case was achieved by generalizing the quasi-Lie bracket
of the canonical case.

The formalisms were presented in such a way
to shed light on
practical implementation via computer simulation algorithms.
The particular class of algorithms upon which we focused is based on
the unfolding of the evolution of the operator-valued quasi-probability function
in terms of piecewise-deterministic trajectories evolving on
the adiabatic energy surfaces of the system.
These~methods scales favorably in terms of bath DOF
but, to date, have been limited
to relatively short time intervals and Markovian systems.
 {When the dynamics is non-Markovian, the memory function,
{i.e.}, the autocorrelation function of the
random force~\cite{bp,gardiner}, cannot be approximated by a delta function.
The memory function of the bath can be expected to become more
and more different from a delta function as the quantum character of the
bath becomes more pronounced (for~example, at~low temperature) and as the
subsystem-bath coupling grows in strength.}

 {The QCLE discussed herein constitutes an
approach to open quantum system dynamics (in the case of hybrid quantum-classical
systems) that is both distinct and complementary to that given
by master equations~\cite{bp,gardiner}. 
Within the QCLE approach, the degrees of freedom of the bath are not
integrated out of the dynamics but are explicitly taken into account
at every time step. Hence, there is
no memory function to be approximated and bath properties
can be calculated with the same ease with which
subsystem properties are computed. The limitations of the QCLE approach are mostly
numerical in character and arise in the SSTP algorithm,
herein discussed, from the momentum-jump approximation and
the accumulation of fluctuating statistical weights associated with the
Monte Carlo sampling of the quantum transitions of the subsystem.}

The QCLE-based approach to quantum dynamics in classical baths
has proven to be successful
in modeling a variety of quantum processes in the condensed 
phase. Nevertheless, the currently algorithms also present significant challenges,
necessitating the need for further improvements and 
developments. In light of the above,
we hope that this review will attract
the attention of a broad community of researchers
and spur further work along this direction.
In addition to further algorithm developments, we are
interested in broadening the scope of applications studied
by this approach. For~example, based on
preliminary results, we believe that this approach
can be successfully applied to studying the interplay
between quantum and classical fluctuations in hybrid 
nanoscale devices.


\vspace{6pt} 



\subsection{* Funding: A.S. and R.G. acknowledge support
by research funds in memory of Francesca Palumbo,
difc 3100050001d08+, University of Palermo.}

\appendix

\section{Representation in the Adiabatic Basis}
\label{sec:adbas_rep}

In the adiabatic basis, Equation~(\ref{eq:qcle}) reads
\begin{equation}
\frac{d}{dt}W_{\alpha\alpha'}(X,t) =
- \sum_{\beta \beta'} i{\cal L}_{\alpha\alpha',\beta\beta'}
W_{\beta\beta'}(X,t) \;,
\label{eq:qcle_adbas}
\end{equation}
where
\begin{eqnarray}
W_{\alpha\alpha'}(X,t)
&=&\langle\alpha; Q|\hat W(X,t) |\alpha'; Q\rangle
\end{eqnarray}
are the matrix elements of the density matrix.
Upon defining the Bohr frequency as 
\begin{equation}
\omega_{\alpha\alpha'}=
\frac{E_{\alpha}-E_{\alpha'}}{\hbar}
\label{eq:Bohr-omega}
\;,
\end{equation}
the Liouville super-operator may be written as
\begin{equation}
i{\cal L}_{\alpha\alpha',\beta\beta'}
=
i\omega_{\alpha\alpha'}\delta_{\alpha\beta}\delta_{\alpha'\beta'}
+
\delta_{\alpha\beta}\delta_{\alpha'\beta'}iL_{\alpha \alpha '}
+
{\cal T}_{\alpha \alpha ',\beta\beta'}\;.
\label{eq:liouville}
\end{equation}
We have also introduced a classical-like Liouville operator
\begin{eqnarray}
iL_{\alpha\alpha '}&=&
\frac{P}{M}\frac{\partial }{\partial Q}
+\frac{1}{2}\left(F_{\rm W}^{\alpha}+F_{\rm W}^{\alpha'}\right)
\frac{\partial}{\partial P}
\;,
\label{eq:classical-like_L}
\end{eqnarray}
where
\begin{eqnarray}
F_{\rm W}^\alpha&=&-\frac{\partial E_{\alpha}}{\partial Q}
\end{eqnarray}
is the Hellmann--Feynman force.

In Equation~(\ref{eq:liouville}), the transition operator
${\cal T}_{\alpha \alpha ',\beta \beta '}$
is defined as
\begin{eqnarray}
{\cal T}_{\alpha\alpha',\beta\beta'}
&=&
\delta_{\alpha'\beta'}
\frac{P}{M}\cdot d_{\alpha\beta}
\left(1+\frac{1}{2}S_{\alpha\beta}
\cdot \frac{\partial }{\partial P}\right)
+
\delta_{\alpha\beta}
\frac{P}{M}\cdot d_{\alpha'\beta'}^*
\left(1 +\frac{1}{2}S_{\alpha'\beta'}^*
\cdot \frac{\partial}{\partial P}\right)
\label{eq:T_aa'bb'}
\;.
\end{eqnarray}

In turn, the transition operator is defined in terms of the
shift vector
\begin{eqnarray}
S_{\alpha\alpha'}
&=&
\frac{\left(E_{\alpha}-E_{\alpha'} \right)}
{\frac{P}{M}\cdot d_{\alpha\alpha'}}d_{\alpha\beta}
\label{eq:S_aa'}
\end{eqnarray}
and of the coupling vector
\begin{eqnarray}
d_{\alpha\alpha'}&=&\langle\alpha; Q|\frac{\partial}{\partial Q}|\alpha'; Q\rangle \;.
\end{eqnarray}


\section{The Nos\'e--Hoover Thermostat}
\label{app:nh}

The Nos\'e--Hoover thermostat was originally formulated 
in Refs.~\cite{nose,hoover}.
Herein, we follow Refs.~\cite{b1,b2,aspvg}.
The Hamiltonian of the subsystem with phase space coordinates
$(R,P)$ is:
\begin{eqnarray}
H^{\rm B}&=& \frac{P^2}{2M}+V(R)\;,
\end{eqnarray}
where $V(R)$ is the potential energy. 
One can introduce an extended system comprised by the coordinates
of the original subsystem augmented with the additional
variables $Q_\eta$ and conjugate momentum $P_{\eta}$.
The dimension of such an extended phase space is obviously
$2N+2$, which is computationally tractable whenever $N$ is
computationally tractable. As a consequence, the  phase space
point of the extended system is
\begin{eqnarray}
X^{\rm N}&=&\left[\begin{array}{c} R\\ Q_\eta \\ P \\
P_{\eta}\end{array}\right]\;,
\end{eqnarray}
while the energy reads: 
\begin{eqnarray}
H^{\rm N}&=&H^{\rm B}+3Nk_BTQ_\eta+\frac{P_{\eta}^2}{2M_{\eta}}\;,
\label{eq:ham-nose}
\end{eqnarray}
where $M_{\eta}$ is a fictitious mass associated with the 
additional degree of freedom, $k_B$ is Boltzmann constant,
and $T$ the bath constant temperature.
In order to define time evolution, we abandon the 
Hamiltonian structure of the theory. To this end, using 
the general formalism of Refs.~\cite{b1,b2,aspvg}, we~introduce
the antisymmetric matrix:
\begin{eqnarray}
\mbox{\boldmath$\cal B$}^{\rm N}=
\left[\begin{array}{cccc} 0 & 0 & 1 & 0\\ 0 & 0 & 0 & 1\\
-1 & 0 & 0 & -P\\ 0 & -1 & P & 0 \end{array}\right]\;,
\label{eq:BN_matrix}
\end{eqnarray}
so that Nos\'e's equations of motion can be written as
\begin{eqnarray}
\dot{X}_K^{\rm N}
&=&
\sum_{I,J=1}^{2(N+1)}
X_K^{\rm N}\overleftarrow{\nabla}_I^{\rm N}
{\cal B}_{IJ}^{\rm N}
\overrightarrow{\nabla}_J^{\rm N} H^{\rm N}
=\sum_{J=1}^{2N}B_{KJ}^{\rm N}
\overrightarrow{\nabla}_J^{\rm N}H^{\rm N}\;,
\label{eq:nosemd}
\end{eqnarray}
where the first equality on the right-hand side of
Equations~(\ref{eq:nosemd}) introduces the Nos\'e bracket,
while~the extended phase space gradient is denoted
as $\nabla_J^{\rm N}=\partial/\partial X_J^{\rm N}$.
We remark here that the Nos\'e bracket does not satisfy
the Jacobi relation~\cite{b1,b2,aspvg},
and thus defines a quasi-Hamiltonian algebra.
The Liouville equation for the Nos\'e distribution function is
\begin{eqnarray}
\frac{\partial}{\partial t}W^{\rm N}(X^{\rm N},t)
&=&
-\sum_{K=1}^{2(N+1)}\nabla_K^{\rm N}
\left(\dot{X}_K^{\rm N}W^{\rm N}(X^{\rm N},t)\right)
\nonumber\\
&=&
-\left(\sum_{K=1}^{2(N+1)}\dot{X}_K
\overrightarrow{\nabla}_K^{\rm N}
-\kappa^{\rm N}\right)W^{\rm N}(X^{\rm N},t)=0\;,
\end{eqnarray}
where the compressibility of the phase space reads:
\begin{eqnarray}
\kappa^{\rm N}
&=&
\sum_{k=1}^{2(N+1)} \nabla_K^{\rm N} \dot{X}_k
=
\sum_{k,j=1}^{2(N+1)}
B_{KJ}^{\rm N} \overleftarrow{\nabla}_K^{\rm N}
\overrightarrow{\nabla}_J^{\rm N} H^{\rm N}
\;.\label{eq:kappanose}
\end{eqnarray}

As implied by Equation~(\ref{eq:kappanose}), Nos\'e's phase space flow 
has a non-zero compressibility (however, this does not always
occur for a quasi-Hamiltonian dynamics).
In terms of the Nos\'e bracket,
the~equilibrium Liouville equation for Nos\'e distribution function reads:
\begin{eqnarray}
W^{\rm N}(X^{\rm N})
\overleftarrow{\nabla}^{\rm N}
\mbox{\boldmath$\cal B$}^{\rm N}
\overrightarrow{\nabla}^{\rm N}
H^{\rm N}
&=&
-\kappa^{\rm N}W^{\rm N}(X^{\rm N})\;.
\label{eq:noseliouvilleeq}
\end{eqnarray}

By direct substitution, one can verify that the solution  of Equation~(\ref{eq:noseliouvilleeq}) is:
\begin{eqnarray}
W^{\rm N}(X^{\rm N})&\propto&\exp\left[-w\right]
\delta(E-H^{\rm N})\;,
\end{eqnarray}
where $w$ is defined by the equation $dw/dt=\kappa^{\rm N}$.
Equations~(\ref{eq:nosemd}) can be written explicitly in the form:
\begin{eqnarray}
\dot{R} &=& \frac{P}{M},\\
\dot{P} &=& -\frac{\partial V}{\partial R}
-P\frac{P_{\eta}}{M_{\eta}},\\
\dot Q_\eta&=&\frac{P_{\eta}}{M_{\eta}},\\
\dot{P}_{\eta}&=&\frac{P^2}{M}-Nk_BT\;.
\end{eqnarray}

In order to write explicitly the Nos\'e distribution function, 
it is useful to introduce the following extended phase space 
function:
\begin{equation}
H^{\rm T}=H^{\rm B}+\frac{P_{\eta}^2}{2M_{\eta}}\;.
\end{equation}

Using the equations of motion, one finds
\begin{equation}
\frac{dH^{\rm T}}{dt}=-Nk_BT\frac{P_{\eta}}{M_{\eta}}\;,
\end{equation}
which is related to the compressibility by
\begin{equation}
\kappa^{\rm N}=-N\frac{P_{\eta}}{M_{\eta}}
=\beta\frac{d H^{\rm T}}{dt}\;.
\end{equation}

At this point, we have all the ingredients that are needed
to prove that extended phase space averages 
of functions of the subsystem coordinates $(R,P)$
can be written as canonical averages. We~start by considering
\begin{eqnarray}
\langle A(R,P)\rangle_{\rm N}
&\propto&\int dX^{\rm N}e^{-\int\kappa^{\rm N} dt}
\delta(E-H^{\rm N}) A(R,P)
\nonumber\\ &&\nonumber\\ &=&
\int dRdPdQ_\eta dP_{\eta}
e^{-\beta\int\frac{dH^{\rm T}}{dt}dt} 
\delta(E-H^{\rm N})A(R,P)\nonumber\\
&&\\
&=& \int dRdPdQ_\eta dP_{\eta}
e^{-\beta H^{\rm T}} \delta(E-H^{\rm N})A(R,P)\;.\nonumber
\end{eqnarray}

The integral
\begin{equation}
\int dQ_\eta\delta(E-H^{\rm N})
\end{equation}
is calculated by using the identity
\begin{eqnarray}
\delta(f(Q_\eta))=
\sum_{\{Q_{\eta_0}\}}\frac{\delta(Q_\eta-Q_{\eta_0})}
{\frac{df}{dQ_\eta}(Q_{\eta_0})}\;,
\end{eqnarray}
where the sum runs over the zeros $Q_{\eta_0}$ of $f(Q_\eta)$.
Upon identifying $f(Q_\eta)=E-H^{\rm N}$, one gets
$Q_{\eta_0}=H^{\rm T}-E/N$ and
\begin{equation}
\delta(f(Q_\eta))
=
\frac{\delta\left(Q_\eta-\beta({\cal H}_T-E)/N\right)}{3Nk_BT}
\;
\end{equation}
with the above results,
the integral over $Q_\eta$ becomes a trivial Gaussian integral
over $P_{\eta}$:
\begin{equation}
\int dP_{\eta}e^{-\beta\frac{P_{\eta}^2}{2M_{\eta}}}
=\sqrt{\pi M_{\eta}k_BT}\;.
\end{equation}

Finally, one obtains: 
\begin{eqnarray}
\langle A(R,P)\rangle_{\rm N}&\propto&
\int dRdP e^{-\beta H^{\rm B}}A(R,P)\equiv
\langle A(R,P)\rangle_{\rm can}\;.
\end{eqnarray}

Hence, averages in the canonical ensemble can be calculated
by letting the trajectories evolve according to Nos\'e's dynamics.

The quasi-Hamiltonian Nos\'e dynamics is a well-established
tool of molecular dynamics simulations.
In practice, it is adopted whenever one wants to calculate
dynamical properties at constant temperature and/or study phase
transitions. Discussions and pointers to the relevant literature
on the subject can be found in Ref.~\cite{frenkel}.


\section{Stationary Operator-Valued Nos\'e Quasi-Probability Function}
\label{app:stationary}

The quantum average of any operator $\hat W^{\rm N}(X^{\rm N})$,
in a dynamics where the temperature of the $X$ degrees of
freedom is controlled by the Nos\'e--Hoover thermostat
can be calculated as
\begin{equation}
\langle \hat{\chi}(X^{\rm N},t)\rangle
={\rm Tr}'\int dX^{\rm N}~\hat W^{\rm N}(X^{\rm N},t)
\hat{\chi}(X^{\rm N}) \;.
\end{equation}

The action of $\exp\left(i{\cal L}^{\rm N} t\right)$ can be transferred
from $\hat\chi(X^{\rm N})$ to
$\hat W^{\rm N}(X^{\rm N})$ by using the cyclic invariance 
of the trace and integrating by parts the terms coming from
the  classical brackets.
One can write
\begin{equation}
i{\cal L}^{\rm N}=\frac{i}{\hbar}
\left[\hat H^{\rm N},\dots \right]-\frac{1}{2}
\hat H^{\rm N}\overleftarrow{\nabla}^{\rm N}
\mbox{\boldmath$\cal B$}\overrightarrow{\nabla}^{\rm N}
-
\overleftarrow{\nabla}^{\rm N}\mbox{\boldmath$\cal B$}
\overrightarrow{\nabla}^{\rm N} \hat H^{\rm N}\}\;.
\end{equation}

The action of $i{\cal L}^{\rm N}$ on an arbitrary operator
$\hat\chi(X^{\rm N})$ is defined by
\begin{eqnarray}
i{\cal L}^{\rm N}\hat\chi
&=&
=\frac{i}{\hbar}
\left[\hat H^{\rm N},\hat\chi \right]-\frac{1}{2}
\hat H^{\rm N}\overleftarrow{\nabla}^{\rm N}
\mbox{\boldmath$\cal B$}\overrightarrow{\nabla}^{\rm N}\hat\chi
-\hat\chi
\overleftarrow{\nabla}^{\rm N}\mbox{\boldmath$\cal B$}
\overrightarrow{\nabla}^{\rm N} \hat H^{\rm N}\;
\end{eqnarray}
when integrating by parts the right-hand side,
one obtains a term
proportional to the compressibility
$\kappa^{\rm N}= \overrightarrow{\nabla}^{\rm N}
\mbox{\boldmath$\cal B$}^{\rm N}
\overrightarrow{\nabla}^{\rm N}\hat H^{\rm N}$.
As a result,
the quantum Liouville operator, partially depending on
phase space variables, is non-Hermitian
\begin{equation}
\left(i\hat{\cal L}^{\rm N}\right)^{\dag} 
=-i\hat{\cal L}^{\rm N}-\kappa^{\rm N}\;.
\end{equation}

The average value can then be written as
\begin{equation}
\langle \hat{\chi}\rangle
={\rm Tr}'\int dX~\hat\chi(X^{\rm N})
\exp\left[-(i{\cal L}^{\rm N}+\kappa^{\rm N}) t\right]
\hat W^{\rm N}(X^{\rm N})\;.
\end{equation}

The operator-valued Nos\'e quasi-probability function
evolves under the equation:
\begin{small} 
\begin{eqnarray} \begin{array}{lll}
\frac{\partial}{\partial t}\hat W^{\rm N}(X^{\rm N},t)
&=&
-\frac{i}{\hbar}
\left[\hat H^{\rm N},\hat W^{\rm N}(X^{\rm N},t)\right]
+\frac{1}{2}
\left(
H^{\rm N}
\overleftarrow{\nabla}^{\rm N}
\mbox{\boldmath$\cal B$}^{\rm N}
\overrightarrow{\nabla}^{\rm N}
\hat W^{\rm N}(X^{\rm N},t)
-
\hat W^{\rm N}(X^{\rm N},t)
\overleftarrow{\nabla}^{\rm N}
\mbox{\boldmath$\cal B$}^{\rm N}
\overrightarrow{\nabla}^{\rm N}
\hat{H}^{\rm N}
\right)
 \\
&-&\kappa^{\rm N}(X)\hat W_{\rm N}(X,t)\;. \end{array} 
\label{eq:qc-dme}
\end{eqnarray}
\end{small}


The stationary operator-valued Nos\'e quasi-probability function
$\hat W^{\rm N,e}$ is defined by
\begin{equation}
(i{\cal L}^{\rm N}+\kappa^{\rm N})\hat W^{\rm N,e}=0\;.
\label{eq:qm-case}
\end{equation}

To find the explicit expression, one can follow Ref.~\cite{qc-stat}:
the density matrix is expanded in powers of $\hbar$
\begin{equation}
\hat W^{\rm N,e}=
\sum_{k=0}^{\infty}\hbar^n\hat W^{{\rm N,e},(k)}
\;
\end{equation}
and an explicit solution in the adiabatic basis is searched for.
On such a basis, the Nos\'e--Liouville operator is expressed by
Equation~(\ref{eq:nose-liouville}) and the Nos\'e Hamiltonian is given by
\begin{eqnarray}
H_{\rm N}^{\alpha}&=&\frac{P^2}{2M}+\frac{P_{\eta}^2}{2M_{\eta}}
+Nk_BTQ_\eta+E_{\alpha}(R)\nonumber \\
&=&H^P_{\alpha}(R,P)+\frac{P_\eta^2}{2M_{\eta}}+Nk_BTQ_\eta
\;.
\end{eqnarray}

One obtains an infinite set of equations corresponding to
the various power of $\hbar$ 
\begin{eqnarray}
iH_{\alpha\alpha'}^{\rm N}W_{\alpha\alpha'}^{\rm N,e(0)}
&=&0
\label{stat-n=0}
\\
iH_{\alpha\alpha'}^{\rm N}W_{\alpha\alpha'}^{{\rm N,e},(k+1)}
&=&
-(iL_{\alpha\alpha'}^{\rm N}+\kappa^{\rm N})
W_{\alpha\alpha'}^{{\rm N,e},(k)}
+\sum_{\beta\beta'}{\cal T}_{\alpha\alpha',\beta\beta'}
W_{\beta\beta'}^{{\rm N,e},(k)}
~ (k\ge 1)\;.
\label{stat-n+1}
\end{eqnarray}

In order to ensure that a solution can be found
by recursion, one must discuss the solution of Equation~(\ref{stat-n+1})
when calculating the diagonal elements $W_{{\rm N}e}^{(n)\alpha\alpha}$ in terms
of the off-diagonal ones $W_{{\rm N}e}^{(n)\alpha\alpha'}$.
To this end, using
$W_{\alpha\alpha'}^{{\rm N,e}(k)}
=(W_{\alpha'\alpha}^{{\rm N,e},(k)})^*$,
${\cal T}_{\alpha\alpha,\beta\beta'}={\cal T}_{\alpha\alpha,\beta'\beta}^*$
and the fact that ${\cal T}_{\alpha\alpha,\beta\beta}=0$ when
a real basis is chosen,
it is useful to re-write Equation~(\ref{stat-n+1}) in the form
\begin{equation}
(iL_{\alpha\alpha}^{\rm N}+\kappa^{\rm N})
W_{\alpha\alpha}^{{\rm N}e,(k)}
=\sum_{\beta>\beta'}2{\rm Re}
\left({\cal T}_{\alpha\alpha,\beta\beta'}
W_{\beta\beta'}^{{\rm N,e}(k)}\right)\;.
\label{stat-condition}
\end{equation}

One has~\cite{b1}
$(-iL_{\alpha\alpha}^{\rm N}-\kappa^{\rm N})^{\dag}
=iL_{\alpha\alpha}^{\rm N}$. The right-hand side of this equation
is expressed by means
of the generalized bracket in Equation~(\ref{gen-bracket}):
$H_{\rm N}^{\alpha}$ and any general function $f(H_{\rm N}^{\alpha})$
are constants of motion under the action 
of $iL_{\alpha\alpha}^{\rm N}$.
The phase space compressibility $\kappa^{\rm N}$ associated
with the generalized bracket in the case of Nos\'e dynamics is
\begin{eqnarray}
\kappa_{\alpha}^{\rm N}
&=&-\beta\frac{d}{dt}
\left(\frac{P^2}{2M}+\frac{P_{\eta}^2}{2M_{\eta}}
+E_{\alpha}(R)\right)
\nonumber \\
&=&-\beta N\frac{P_{\eta}}{M_{\eta}}
=-\beta N\frac{d}{dt}H_{\alpha}^{\rm T}\;,
\end{eqnarray}
where $ N$ is the number of classical momenta $P$
in the Hamiltonian.

To ensure that a solution to Equation~(\ref{stat-condition})
exists, one must invoke the theorem
of Fredholm alternative, requiring that the 
right-hand side of Equation~(\ref{stat-condition}) 
is orthogonal to the null space of 
$(iL_{\alpha\alpha}^{\rm N})^{\dagger}=-iL_{\alpha\alpha}^{\rm N}
-\kappa^{\rm N}$~\cite{hilbert}.
The null-space of this operator
is defined by the equation \linebreak
$(iL_{\alpha\alpha}^{\rm N}+\kappa^{\rm N})G_\alpha(X)=0$,
with $G_\alpha(X)=
f(H_{\alpha}^{\rm N}) \exp(-w_{\alpha}^{\rm N})$.
Hence, the condition to be satisfied is
\begin{equation}
\int dX^{\rm N}e^{-w_\alpha}
\sum_{\beta>\beta'}2{\rm Re}
\left({\cal T}_{\alpha\alpha,\beta\beta'}
W_{\beta\beta'}^{{\rm N,e},(k)}\right)
f(H_{\alpha}^{\rm N})=0 \;.
\label{fredholm}
\end{equation}

The fact that
$2\exp(-w_\alpha){\rm Re}\left({\cal T}_{\alpha\alpha,\beta\beta'}
W_{\beta\beta'}^{{\rm N,e},(k)} \right)$
and $f(H_{\rm N}^{\alpha})$ are, respectively,
an odd and an even function of $P$
guarantees the validity of Equation~(\ref{fredholm}).

The formal solution of Equation~(\ref{stat-condition}) can then be written as
\begin{equation}
W_{\alpha\alpha}^{{\rm N,e},(k)}
=(iL_{\alpha\alpha}^{\rm N}+\kappa^{\rm N})^{-1}
\sum_{\beta>\beta'}2{\rm  Re}
\left({\cal T}_{\alpha\alpha,\beta\beta'}
W_{\beta\beta'}^{{\rm N,e},(k)} \right)\;,
\label{eq:sol1}
\end{equation}
and the formal solution of Equation~(\ref{stat-n+1}) for 
$\alpha\neq\alpha'$ as
\begin{eqnarray}
W_{\alpha\alpha'}^{{\rm N,e},(n+1)}
&=&
\frac{i}{E_{\alpha\alpha'}}
(iL_{\alpha\alpha'}^{\rm N}+\kappa^{\rm N})
W_{\alpha\alpha'}^{{\rm N,e},(k)}
-\frac{i}{H_{\alpha\alpha'}^{\rm N}}
\sum_{\beta\beta'}
{\cal T}_{\alpha\alpha',\beta\beta'}
W_{\beta\beta'}^{{\rm N,e},(k)} \;.
\label{eq:sol2}
\end{eqnarray}

Equations~(\ref{eq:sol1}) and~(\ref{eq:sol2}) allow one to calculate
$W_{\alpha\alpha'}^{\rm N,e}$ to all orders in $\hbar$
once $W_{\alpha\alpha'}^{{\rm N,e},(0)}$ is given.
This order zero term is obtained by the solution of
$(iL_{\alpha\alpha}^{\rm N}+\kappa^{\rm N})
W_{\alpha\alpha}^{{\rm N,e},(0)}=0$. All higher order terms
are obtained by the action of
$H_{\alpha\alpha'}^{\rm N}$, the imaginary unit 
$i$ and ${\cal T}_{\alpha\alpha'\beta\beta'}$ (involving factors of 
$d_{\alpha\alpha'}$, $P$ and derivatives with respect to $P$.
Hence, one can conclude that functional dependence of
$W_{\rm Ne}^{(0)\alpha\alpha}$ on
the Nos\'e variables $Q_\eta$ and $P_{\eta}$
is preserved in higher order terms
$W_{\alpha\alpha'}^{{\rm N,e},(n)}$.
One can find a stationary solution to order $\hbar$
by considering the first two equations of the set
given by Equations~(\ref{stat-n=0}) and~(\ref{stat-n+1}):
\begin{eqnarray}
\left[\hat{H}^{\rm N},\hat W^{{\rm N,e},(0)}\right]& =&0  
\qquad ({\rm for~}k=0)\;,
\label{n=0}
\\
i\left[\hat H^{\rm N},\hat W^{{\rm N,e},(1)}\right]&=&
+\frac{1}{2}\left(\hat{H}^{\rm N}
\overleftarrow{\nabla}{\cal B}^{\rm N}
\overrightarrow{\nabla}
\hat W^{{\rm N,e},(0)}
+ \hat W^{{\rm N,e},(0)}
\overleftarrow{\nabla}{\cal B}^{\rm N}
\overrightarrow{\nabla}
\hat{H}^{\rm N}\right)
~({\rm for~}k=1)\;.
\label{n=1}
\end{eqnarray}

For the $\hbar^0$ term, one can make the ansatz
\begin{equation}
\hat W_{\alpha\beta}^{{\rm N,e},(0)}
=\frac{1}{Z^{\rm N}}e^{w_\alpha^{\rm N}}
\delta\left({\cal E}_\alpha-H_\alpha^{\rm N}\right)\delta_{\alpha\beta} \;,
\label{ansatz}
\end{equation}
where $Z^{\rm N}$ is
\begin{eqnarray}
Z^{\rm N}&=&\sum_{\alpha}\int d{\cal M}~
\delta\left({\cal E}_\alpha-H_\alpha^{\rm N}\right)
\end{eqnarray}
and obtain
\begin{eqnarray}
\hat W_{\alpha\alpha'}^{{\rm N,e},(1)}
&=&-i
\frac{P}{M}d_{\alpha\alpha'}\hat W_{\alpha\alpha}^{{\rm N,e},(0)}
\left[\frac{1-e^{-\beta(E_{\alpha'}-E_{\alpha})}}
{E_{\alpha}-E_{\alpha'}}
+\frac{\beta}{2}
\left(1+e^{-\beta(E_{\alpha'}-E_{\alpha})}\right)
\right]
\label{ansatz2}
\end{eqnarray}
for the $\hbar$ term.

Equations~(\ref{ansatz}) and~(\ref{ansatz2}) give the explicit form
of the stationary solution of the Nos\'e-Liouville equation
up to order ${\cal O}(\hbar)$. 
One can now prove that,
when calculating averages of quantum-classical operators
depending only on physical phase space variables,
${\cal G}_{\alpha}(R,P)$,
the canonical form of the stationary density is obtained.
It can be noted that it will suffice to prove this result
for the $\hbar^0$ term since,
as discussed before, the differences with
the standard case are  contained  therein.
Indeed,~when~calculating
\begin{eqnarray}
\langle{\cal G}_{\alpha}(R,P)\rangle
&\propto&=\sum_{\alpha}\int dX^{\rm N}e^{-w_\alpha^{\rm N}}
{\cal G}_{\alpha}(R,P)
\delta({\cal E}_\alpha-H_\alpha^{\rm T}-Nk_{B}TQ_\eta)\;,
\end{eqnarray}
considering the integral of the delta function over
Nos\'e variables, one has
\begin{eqnarray}
\int dP_{\eta}dQ_\eta~e^{-N\eta}
\delta({\cal E}_\alpha-H_{\alpha}^{\rm T}-Nk_{B}TQ_\eta)
={\rm const}\times
\exp[-\beta H_{\alpha}^{\rm T}]\;,
\nonumber \\
\end{eqnarray}
where the property
$\delta(f(s))=[df/ds]^{-1}_{s=s_0}\delta(s-s_0)$ has been used
($s_0$ is the zero of $f(s)$).


\end{document}